\documentclass[]{rmaa}
\usepackage[dvips]{epsfig}
\usepackage{paralist}
\usepackage[latin1]{inputenc}
\title{Physical effects of gas envelopes with different extension on
the collapse of a gas core.}
\altaffiltext{1}{Centro de Investigaci\'on en F\'{\i}sica de la
Universidad de Sonora, Mexico.}

\author{Guillermo Arreaga-Garc\'ia\altaffilmark{1} and Julio Saucedo Morales,\altaffilmark{1}}

\fulladdresses{\item Centro de
Investigaci\'on en F\'{\i}sica de la
Universidad de Sonora, M\'exico, Apdo. Postal 14740, C.P. 83000,
Hermosillo, Sonora, Mexico. (garreaga@cajeme.cifus.uson.mx)}
\resumen{En este trabajo estudiamos el colapso gravitacional de una
nube de gas de hidr\'ogeno molecular compuesta de
un n\'ucleo m\'as un envolvente
de gas rodeando al n\'ucleo. Simulamos num\'ericamente el colapso de cuatro
modelos de nube para entrever la evoluci\'on temporal de algunas
variables din\'amicas, entre otras, el momento angular y la raz\'on $aem$; las razones
entre las energias t\'ermica y rotacional con respecto a la energ\'{\i}a
potencial gravitacional, denotadas como $\alpha$ y $\beta$, respectivamente. 
Re-tomamos los modelos introducidos 
por ~\citet{NuestroAA}, para hacer una
caracterizaci\'on cuantitativa de los diferentes resultados del colapso
de la nube por medio de las variables din\'amicas ya mencionadas.
Mostramos que podemos comparar
cuantitativamente los efectos de la extensi\'on del envolvente de gas
sobre el colapso del n\'ucleo.}
\abstract{ In this paper we study the gravitational collapse of a
molecular hydrogen gas cloud composed of a core plus a gas envelope
surrounding the core. We numerically simulate the collapse of four cloud
models to take a glimpse to the time evolution of several dynamic variables, such
as the angular momentum and the $aem$ ratio, as well as the ratios between the
thermal and rotational energies with respect to the potential gravitational
energy, denoted as $\alpha$ and $\beta$, respectively, among others.  
We re-take those models introduced by ~\citet{NuestroAA} in
the present paper in order to produce different outcomes of the collapsing cloud
characterized in terms of the aforementioned dynamical
variables. Such characterization was missing in the paper by ~\citet{NuestroAA}, and
here we show that the gas envelope extension effects on the collapsing core can
be quantitatively compared.}
\shortauthor{Arreaga et al.}
\shorttitle{dynamics of collapsing...}
\SetYear{2011}
\addkeyword{dynamical variables}
\addkeyword{integral properties}
\addkeyword{collapse}
\begin{document}
\maketitle
\section{Introduction}
\label{sec:intro}

It is suggested by observational evidence that most young stars in the
Galaxy (around $50$\%) are coupled in binary systems. Astronomical observations
and theoretical studies point out to the clouds early fragmentation as the leading
mechanism to explain the binary stellar systems origin and properties (see
the review by ~\citet{bodeniv}, where several theoretical mechanisms for
binary formation are discussed). Thus, fragmentation is a very important
physical phenomenon whose occurrence in clouds is paramount. Indeed, cloud
fragmentation is one of the key physical events that any plausible
theory of star formation  must include, in order to explain the observed
property that most new born stars are clustered in binary or multiple groups.

Recently, ~\citet{NuestroAA} have studied the protostellar clouds gravitational
collapse, including rotation, thermal pressure and a
centrally condensed radial density profile. They considered a cloud model
composed of a core plus a gas envelope surrounding it; they reported
the outcome of a cloud model depending on both the extension and
mass of the gas envelope. The novel idea that these authors 
worked out was the change of the radial extension of the outer envelope.

In this paper we focus on trying to quantitatively characterize
the most important dynamical events of the full three dimensional set
of numerical hydrodynamical simulations introduced by ~\citet{NuestroAA}.  We 
consider again those cloud models to achieve such a goal, focusing now 
on some dynamical
variables, mainly on the acceleration, density, specific angular
momentum $J/M$ and the $aem$ ratio\footnote{Where $c$ is the speed
of light and $G$ is the Newton gravitational constant.}
$aem \equiv \frac{c \, J}{G \, M^2}$, as well as in the ratios between the thermal and
rotational energies with respect to the gravitational potential
energy, denoted by $\alpha$ and $\beta$, respectively. We prove 
here that these dynamic variables characterize rather well the
collapse process by capturing the most representative physical events,
including fragmentation, which leaves an imprint on these dynamical 
variables. It must be noted that the $aem$ ratio is a dimensionless 
measure of the specific angular momentum, which fortunately has been 
estimated for protostellar clouds. For instance, \citet{fel} reported 
that protostellar cores around one solar mass have an $aem$ ratio 
of about $10^5$. We use here this $aem$ ratio for measuring the 
redistribution of mass and angular momentum during the cloud collapse, 
particularly in its central region. The general result obtained is the 
systematic decrease of the $aem$ ratio  for the particles 
involved in the central cloud collapse.

It must be emphasized that no plots with integral or mechanical properties
of the cloud models were reported in the paper by \citet{NuestroAA}. This
absence is now the main concern for the present work, which must be
considered as an extension of the work by \citet{NuestroAA}, as we now study
how the different collapsing outcomes are manifested on the dynamic
variables. Consequently, we include several plots to envisage the cloud
mechanical state as the gravitational collapse takes 
place.  Hence, we expect that the physical characterization of the 
models, and above all, the integral properties calculation, will be 
useful for a better understanding of the physics of the 
gravitational collapse.

Let us recall that there are several papers demonstrating
that the fate of a collapsing cloud is ultimately determined by the
initial values of its $\alpha$ and $\beta$ energy ratios, see for
instance \citet{toru} and references therein. Thus, for allowing
a comparison to be made between our models and those of other authors, we
have set fixed values chosen from the collapse literature to 
these ratios for all our cloud models. Furthermore, since all the clouds
considered in the paper hereby have the same radial density profile, we
find ourselves in a good position to study the different gas envelopes effects
on the collapse of a unique central gas core, as all the models share the
same innermost central density region.

We mention that one of the most relevant effects of the envelopes on
the collapsing core is the different extension of the spiral arms that
develop around the densest clump, that forms at the cloud central region. Thus,
the results of our models provide a representative sample of scenarios composed
by a central clump plus spiral arms of different lengths.

Recently, ~\citet{tsukamoto} carried out a large set of simulations whose
outcomes span a wide sample of stellar systems formed by a
central protostar plus a surrounding gas disk.  Indeed, our sample is
a subset of reported by ~\citet{tsukamoto}, but in our case, we
only need to deal with a single value for each of the $\alpha$ and
$\beta$ energy ratios, whereas these authors used a wide range of
pairs of values to characterize many independent clouds. Therefore, we
can consider that we have the same cloud collapsing several times
to produce different outcomes, which can be attributed mainly to the
different extension of the gas envelope, this means we only need
to smoothly change the cloud conditions in order to see new physical
scenarios resulting from the collapse.

Moreover, the nature of the disk-star system can have a very deep
influence on the star formation process by means of the disk interaction
with the central star through gravitational torques, as it was
demonstrated by ~\citet{lin}. One of the most interesting results that
these authors found was that the central protostar spin rate is remarkably
reduced by the action of gravitational forces upon it. They argued
that the evolution of the central protostar angular momentum is strongly influenced
by the external fluid surrounding it.

We track the paths of the particles entering into the central clump during
the simulation in the paper hereby, looking forward to see how those
particles lose angular momentum, as a consequence of both the shear
viscosity and the gravitational potential lines rearrangement caused
by the bar-like deformations in the geometry of the densest
central mass distribution. We particularly show how the dynamical
orbital angular momentum and the $aem$ ratio are
assembled from the properties of the individual particles. These orbital
properties are also ultimately regulated by the gas envelope extension.

The outline of the paper is as follows. In 
Section~\ref{sec:inicond} we present in detail the cloud
models of \citet{NuestroAA}, which are studied in this paper too, 
enable a better reading. We
introduce the behavior of the most familiar dynamical variables relevant
to the collapsing cloud in Section \ref{sec:resul}, namely: mass,
density, acceleration, angular momentum and the energy ratios. Later, we discuss
how the different outcomes of the cloud models are recorded on some of
the aforementioned dynamical variables. We try to explain the most relevant
physical events during the collapse, in Section\ref{sec:dis}, in terms of the
dynamical variables behavior describing the cloud evolution. Finally, we
remark the importance that the results of a simulation can eventually be
explained in terms of its dynamical variables in Sect.\ref{sec:con}.
\section{Initial conditions of the suite of cloud models}
\label{sec:inicond}

As we aim to study the gas envelope effects on the core collapse
quantitatively, we consider four cloud models labeled as
$A0$, $A1$, $A2$ and $A3$, each with different extension of the
envelope relative to the core radius, that is,
$R_0/R_c=$ $0.5$, $1.5$, $2.5$ and $3.7$, respectively, where
$R_0$ is the cloud radius and $R_c$ the core radius. We illustrate that
the cloud models under consideration are centrally condensed in the left panels of
Figs.\ref{figdelmodelo} and \ref{funcionplummer};  we summarize 
the values of the most important physical
parameters used for setting up these models in Table \ref{tab:modelos}.
\begin{table}
\caption{The collapse models.}
{\begin{tabular}{|c|c|c|c|c|c|}
\hline
Model & $\frac{R_0}{R_c}$ & $M_0$ & $\rho_0$ & $\Omega_0$ & $c_0$ \\
 & & gr & $gr/cm^3$ & $\times 10^{-13}$ rad/sec & cm/sec  \\
\hline \hline
A0 & 0.5 & $6.28 \times 10^{32}$ & $2.23\, \times 10^{-18}$ & 5.80 & 10620.35 \\
\hline
A1 & 1.5 & $5.14 \times 10^{33}$ & $6.95\, \times 10^{-19}$ & 3.90 & 18751.51 \\
\hline
A2 & 2.5 & $8.34\,\times 10^{33}$& $2.43\, \times 10^{-19}$ & 2.95 & 19942.80  \\
\hline
A3 & 3.7 & $1.04 \times 10^{34}$ & $9.22 \, \times 10^{-20}$& 2.22 & 20344.14  \\
\hline
\end{tabular}
\label{tab:modelos}}
\end{table}

\subsection{The initial radial density profile.}
\label{subsec:radialdensity}

Protostellar collapse models with central condensations were first
studied by \citet{boss87},\citet{boss91} and
subsequently by \citet{siga94} and \citet{siga96}, among others.
In fact, the model of a centrally condensed cloud studied
by ~\citet{NuestroAA} has been called a Plummer cloud, because the
radial density profile used for the cloud was inspired in the
following Plummer-like function

\label{sec:Int}
\begin{equation}
\rho(r)=\rho_c\, \left( \frac{R_c}{\sqrt{r^2+R_c^2}} \right)^\eta \;
, \label{distdens}
\end{equation}
\noindent where we have fixed the free parameters to the following
values:

\begin{equation}
\begin{array}{l}
\rho_c=3.0\times 10^{-18} \; gr \, cm^{-3} = 8.96 \times 10^5 \;
molecules \, cm^{-3} \; , \vspace{0.25 cm} \\
R_c=8.06 \times 10^{16} \; cm = 0.026 \; pc \; , \vspace{0.25 cm} \\
\eta=4 \; .
\end{array}
\label{fparam}
\end{equation}
\noindent as suggested by \citet{whithworth}.

\begin{figure}
\begin{center}
\begin{tabular}{cc}
\includegraphics[width=2.0in, height=2.0 in]{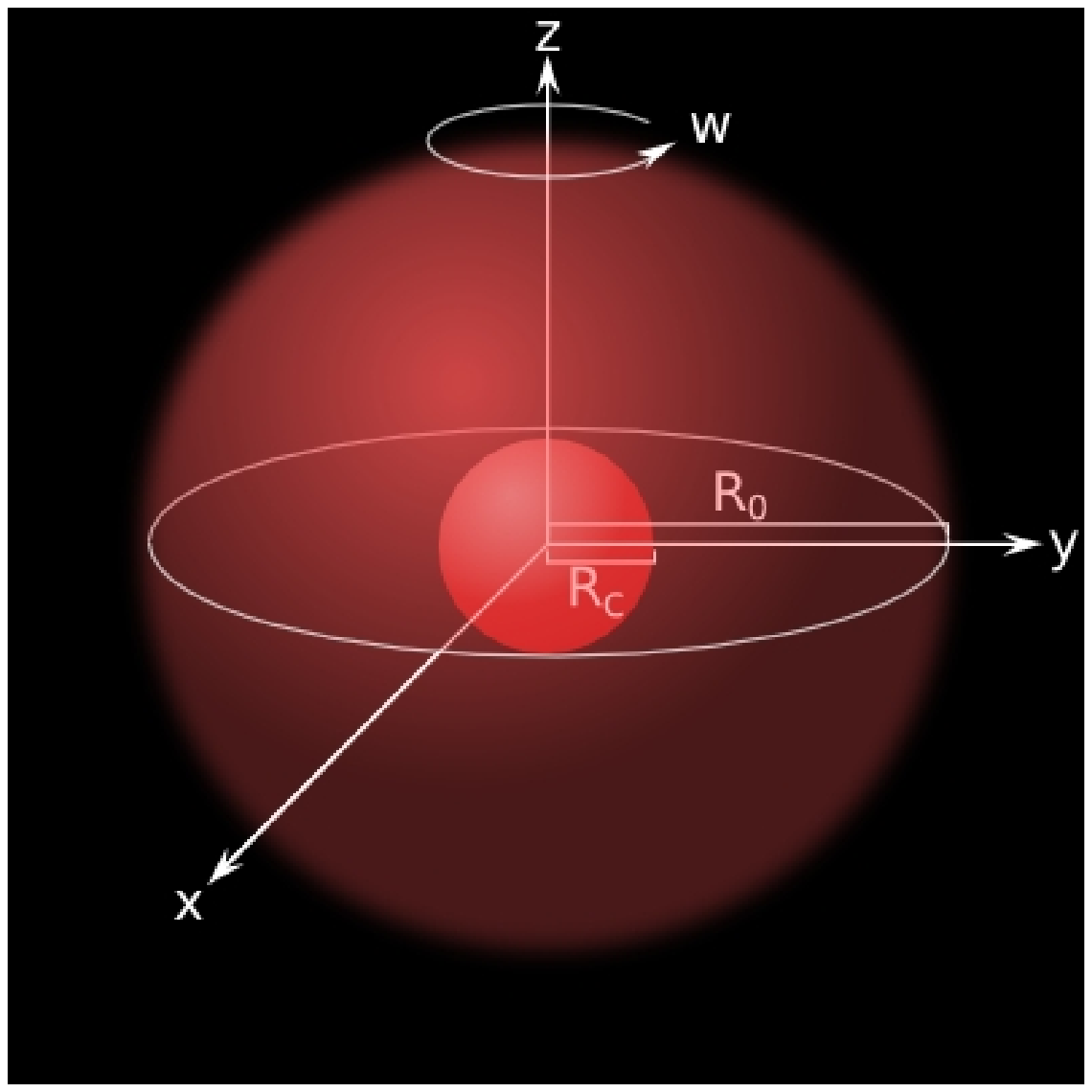}&\includegraphics[width=3.0in,height=2.0 in]{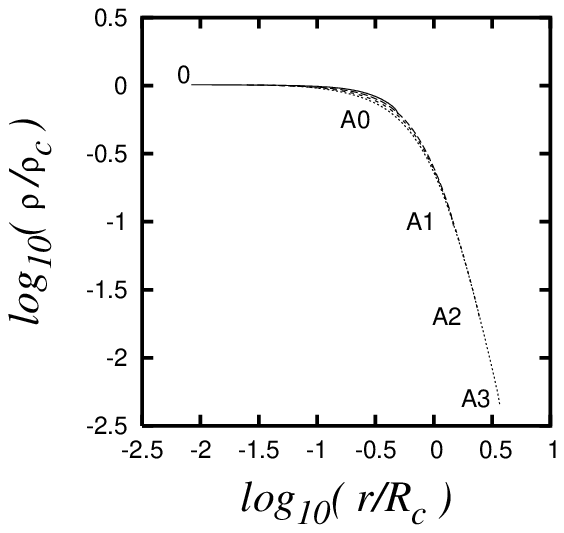}
\end{tabular}
\caption{The model of a cloud composed by a dense core
surrounded by an envelope (left) and the radial density profile measured
for the initial configuration of particles (right). The radial
extension marked in each curve is different according to the model.}
\label{figdelmodelo}
\end{center}
\end{figure}

\begin{figure}
\begin{center}
\begin{tabular}{cc}
\includegraphics[height=2.0 in, width=3.0in]{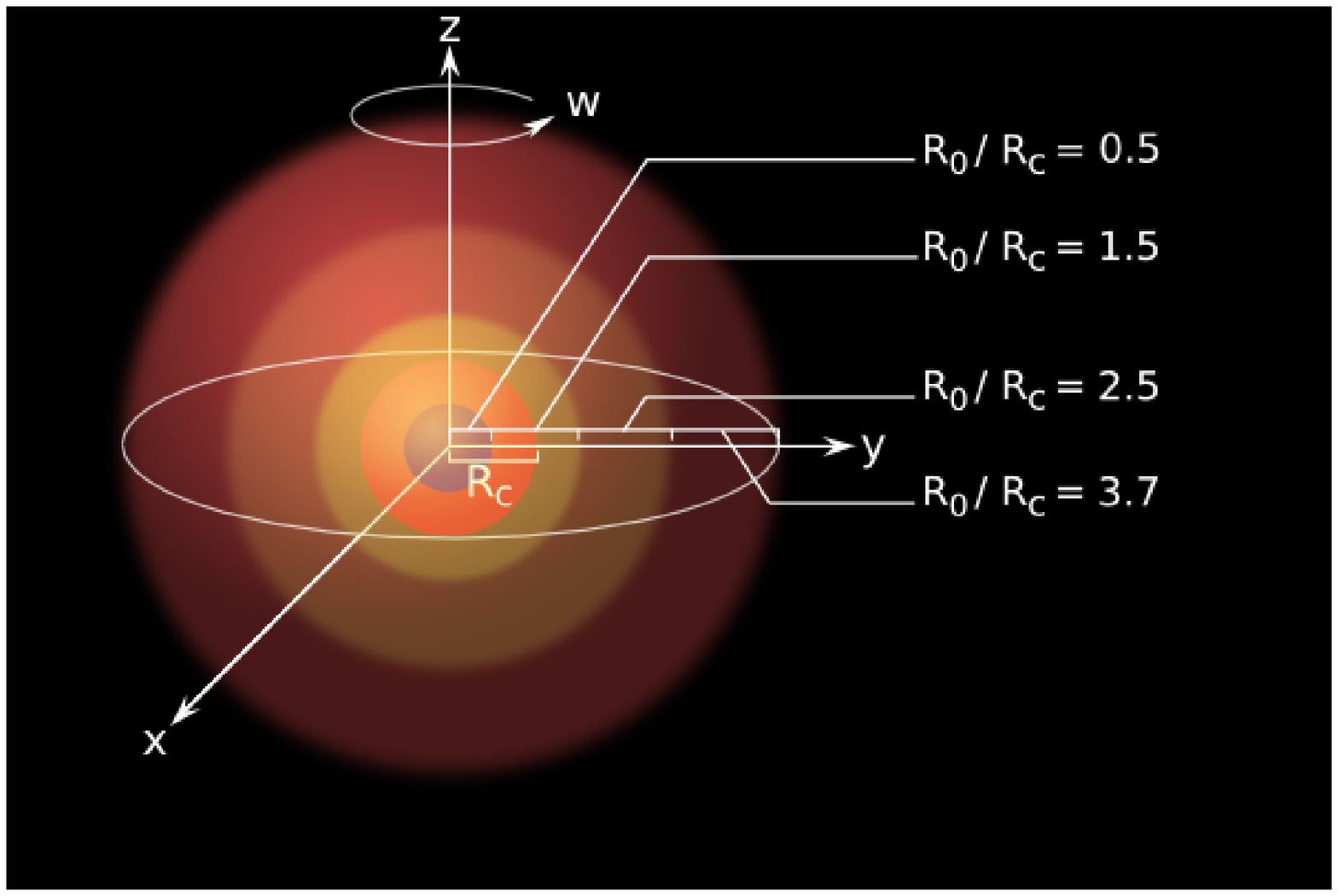}&\includegraphics[height=2.0 in,width=3.0in]{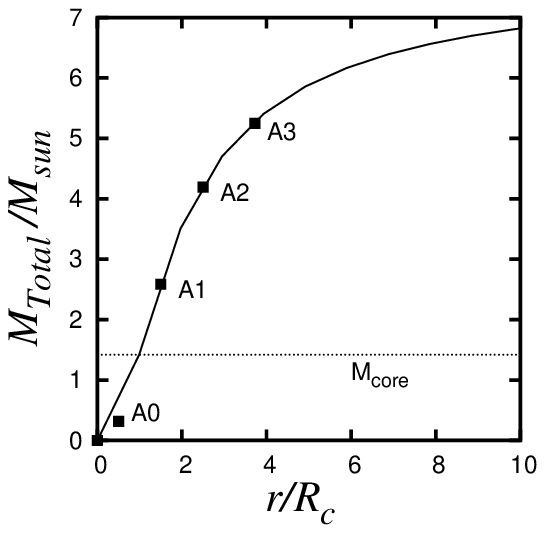}
\end{tabular}
\caption{Schematic representation of the cloud models indicating the
extension of the gas envelopes (left) and the mass contained in the cloud
( from integration of the Plummer
function [solid line] and from the initial
configuration of $SPH$ particles [black squares] )
as a function of radius (right).}
\label{funcionplummer}
\end{center}
\end{figure}

We have added the labels $A0$, $A1$, $A2$ and $A3$ in the plots, to indicate the
cutting radii of the Plummer cloud, whose different extensions define each of
the simulation models, as illustrated in the right panel of
Fig.\ref{funcionplummer}. We mention that the density curves in the right
panel of Fig.\ref{figdelmodelo} show slight and unimportant differences in
the cloud central region (near the $A0$ label). As all the clouds share the
same radial density profile in its innermost parts, then we consider that all the
cloud models share the same core, despite of the fact that the models
differ in other physical parameters, such as the angular velocity.

We also mention that the radial Plummer function shown in Eq.\ref{distdens} does
not "exactly" satisfy  the isothermal Lane-Emden equation, which determines the
solution for an isothermal cloud in an equilibrium configuration. However, what is
most important for us is that the qualitative behavior of the Plummer-like
profile behaves very similar to an approximate analytic solution of the Lane-Emden
equation for the isothermal sphere, as it was found by~\citet{nata}; their approximate
solution is accurate within $0.04 \%$ with the Plummer density profile.

By comparing the Fig. 1 of~\citet{nata}
with the right panel of our Fig.~\ref{figdelmodelo}, we conclude that
these functions behave mathematically almost identically. Moreover, as it was
demonstrated by ~\citet{whithworth}, what becomes more useful after considering
the Plummer profile as a model of protostellar collapse, is that the physical
quantities have simple analytic forms, thus avoiding numerical methods as the only
tool of analysis. Consequently, we have found it to be worthwhile to consider
simulations with the Plummer radial density profile.
\subsection{The initial assembly of particles.}
\label{subsec:iniasembly}

We have accomplished to have a set of $N=10$ million $SPH$ particles
representing the initial cloud configuration with the aforementioned radial density
profile. It should be noticed that the $SPH$ particles do not always have the
same mass $m_i$ in a simulation, for two reasons. The first is that 
each particle mass is
determined by its coordinates location $(x_i,y_i,z_i)$, according to the
density profile, that is, $m_i= \rho(x_i,y_i,z_i)*\Delta x\, \Delta y\, \Delta z$
with $i=1,...,N$, where $\Delta x$ indicates the size of each dimension of the 
rectangular grid in which the particles are initially located. The cloud space volume 
was covered with a total of $286^3$ grid elements. The second reason was 
that a density perturbation was applied
initially by hand to the mass of each $SPH$ simulation
particle $m_i$ in all of the cloud models according to:

\begin{equation}
m_i=m_0 \left (1+a \cos\left(m\, \phi_i \right) \right) \; ,
\label{pertmasa}
\end{equation}
\noindent where $m_0$ is the mass of the $SPH$
simulation particle, and we set the perturbation amplitude to
$a=0.1$, while the mode is fixed to $m=2$. It was done with the purpose of 
favoring a binary protostar development in the cloud innermost region at the 
end of the simulation. These values of $m$ and $a$ have been chosen as it 
is customary in this field of work; see for instance,
\citet{bb93}, \citet{bb96} and \citet{review01}.

We show the total mass contained in the Plummer cloud in the right panel 
of Fig.~\ref{funcionplummer}, which is always an increasing
function of the cloud radius $r$. As an accuracy confirmation of our initial 
particles configuration, the mass calculated from the integration of the 
Plummer function and from the initial configuration of our $SPH$ particles 
agrees very well, as expected.
\subsection{Initial energies.}
\label{sec:inicialener}

The initial cloud for all the models considered in this paper is in counterclockwise
rigid body rotation around the $z$ axis; therefore,
the initial velocity for the $i-th$ $SPH$ particle is given by
$\vec{v}_i= \vec{\Omega_0} \times \vec{r}_i =
(-\Omega_0\, y_i,\Omega_0\, x_i,0)$,
where $\Omega_0$ is the angular velocity magnitude, which has a
different value depending on the cloud model, see Table \ref{tab:modelos}.

It is important to emphasize that all of our cloud models
initially have the same thermal and rotational energy ratio
with regards to the gravitational energy, which are denoted
by $\alpha_0$ and $\beta_0$, respectively\footnote{See Sect.\ref{subs:calproc} for a
detailed definition of $\alpha$ and $\beta$ in the frame of the 
$SPH$ technique.}.  As a matter of
fact, in Table \ref{tab:modelos} we have also reported the initial sound 
speed $c_0$ and the
initial angular velocity $\Omega_0$ given to each cloud model in order to 
have the following numerical ratios:

\begin{equation}
\begin{array}{l}
\alpha_0 \equiv \frac{E_{therm}}{\left|E_{grav}\right|}=0.26 \; , \vspace{0.25 cm}\\
\beta_0 \equiv \frac{E_{rot}}{\left|E_{grav}\right|}=0.16 \; .
\end{array}
\label{defalphabeta}
\end{equation}
\noindent These $\alpha_0$ and $\beta_0$ values were chosen to allow a
direct comparison with other authors, see for example
\citet{bodeniv}. Regarding our models, the $\beta_0 = 0.16$ value gives a
cloud angular velocity $\Omega_0 \, \sim (2.22-5.80) \, \times
10^{-13} \, s^{-1}$; for the case of the uniform density standard
isothermal test, $\alpha_0=0.25$ and $\beta_0=0.20$ that gives
$\Omega_0 \, = \, 1.56 \times 10^{-12} \, s^{-1}$
(see \citet{boss79} and \citet{siga97}), which is an order of
magnitude higher than our $\beta_0$ range values.

We calculate the energy ratios for the core alone in order to
illustrate the energy sharing
mechanism between the core and the envelope, neglecting those particles 
whose distance to
the cloud center is greater than $R_c$; we ignore all the $SPH$ particles whose
radius coordinate $r_i$ is $r_i>R_c$. The results are presented in
Table~\ref{tab:alphasybetascores}, where the $\alpha_c$ and $\beta_c$ values are
calculated up to the core radius $R_c$.
\begin{table}
\caption{Energy ratios calculated up to the core
radius.}
{\begin{tabular}{|c|c|c|c|}
\hline
Model & $\alpha_c$ & $\beta_c$ & $\alpha_c+\beta_c$  \\
\hline \hline
A0 & 0.2643 & 0.1618 & 0.4261  \\
A1 & 0.3002 & 0.1114 & 0.4116  \\
A2 & 0.3350 & 0.0657 & 0.4007  \\
A3 & 0.3547 & 0.0374 & 0.3921  \\
\hline
\end{tabular}
\label{tab:alphasybetascores}}
\end{table}
As expected, according to Table~\ref{tab:alphasybetascores} the core dynamical
properties for the initial configuration in model $A0$ are identical to the whole
cloud dynamical properties, as set by Eq.\ref{defalphabeta}.  Another observation
from Table~\ref{tab:alphasybetascores} is that the larger the gas envelope, the
greater the core thermal energy and, at the same time, the core
rotational energy is smaller. This statement will have important consequences to
explain the  different outcomes derived from the simulations, as it will
be seen in Section~\ref{subs:GlobalProp}.

It is also noteworthy to appreciate by looking at the third column of 
Table~\ref{tab:alphasybetascores}, that
the sum of the energy ratios is always below $0.5$ for all models. This feature is
important as it sets the cloud general tendency to collapse, as dictated by the
virial theorem, which would apply if the hydrogen cloud were in thermodynamical
equilibrium. If this was the case, the energy ratios would satisfy the virial equation.

\begin{equation}
\alpha + \beta=\frac{1}{2}\;,
\label{abvirial}
\end{equation}
\noindent a relation which will be used in some of the plots of
the following sections.
\subsection{The equation of state.}
\label{sec:inicialeos}

Once gravity has produced a substantial contraction of the cloud, the gas begins to
heat. We use a barotropic equation of state in order to take this fact
into account, as it was originally proposed by
\citet{boss2000}, to model the gas thermodynamics:

\begin{equation}
p= c_0^2 \rho \left[ 1 + \left(
\frac{\rho}{\rho_{crit}}\right)^{\gamma -1 } \right] \; ,
\label{beos}
\end{equation}
\noindent where $\rho_{crit}$ defines the critical density above
which the collapse changes from isothermal to adiabatic, and for a
molecular hydrogen gas the ratio of specific heats is $\gamma\, \equiv
5/3$. Furthermore, in this paper we consider only the value

\begin{equation}
\rho_{crit}=5.0\,\times 10^{-14}\; gr/cm^3 \;,
\label{critdensi}
\end{equation}
\noindent chosen to allow a direct comparison with \citet{boss2000}, who calculated a
uniform and gaussian cloud with a barotropic equation of state  
considering the Eddington approximation.
\subsection{Initial angular momentum.}
\label{subs:initmodels}

The importance of studying the
origin of the angular momentum has been
reviewed by \citet{boden95} and \citet{zinne}. In fact, the observed
values of $J/M$ and $aem$ for pre-main sequence stars are lower than those observed for
typical rotating protostellar cores, implying that mass and angular momentum
should be redistributed somehow to ensure a decrease of the $J/M$ and $aem$
by factors of $10^3$ to $10^4$ during star formation, see also \citet{siga94}.

It must also be mentioned
that observations by ~\citet{goodman} have shown that dense cores have velocity
gradients of about $0.3$ to $4.0 \, km \, s^{-1} \, pc^{-1}$, which 
correspond to angular
velocities in the range of $\Omega_0 \sim 9.6 \times 10^{-15} \, s^{-1}$
to $\sim 1.2 \times 10^{-13} \, s^{-1}$, values
which are slightly below ours.

The observational relation between angular momentum and the cloud radius for molecular
cores has been reported by ~\citet{goodman} (see their Fig. 13). Indeed, this plot has
been reproduced and improved by
~\citet{boden95} in his Fig. 1. Let us consider for instance our
model $A3$, with $R_0=0.097 \, pc$; an
associated specific angular momentum  $j=6.3 \times 10^{21} \, cm^2 \, s^{-1}$,
which corresponds to an angular velocity $\Omega=1.75 \times 10^{-13} s^{-1}$; a
value which is very close to ours
for model $A3$. Hence, as suggested by observations, our angular velocity range
is also similar to their numerical simulations.

\begin{figure}
\begin{center}
\includegraphics[width=3.0in]{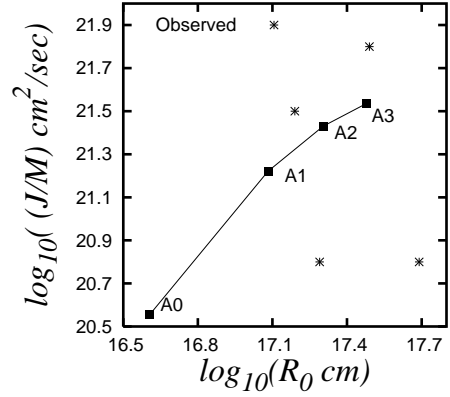}
\caption{The initial specific angular momentum $J/M$ against
the cloud radius $R_0$ for all collapse models. The asterisks mark the values of $J/M$
observed for real clouds, see ~\citet{goodman,boden95}. }
\label{LogJeMvsR0}
\end{center}
\end{figure}

The specific total angular momentum for all the initial configuration of particles
defining our models, is shown in Fig.~\ref{LogJeMvsR0}.
We have introduced asterisks in Fig.\ref{LogJeMvsR0}  to mark the observed quantities to
make clear that the dynamical properties of our clouds are typical when 
compared with the observations reported  by~\citet{goodman} and ~\citet{boden95}.

Curiously, for the initial configuration of model $A0$ there is no
gas envelope, as the cloud extends only to $R_c/2$. Model $A0$ has the highest initial
density and the highest rotational speed because it has the smallest cloud size, as it
can be seen in Table~\ref{tab:modelos}. Besides, its specific angular 
momentum is lower than those already observed for protostellar 
clouds, see Fig.~\ref{LogJeMvsR0}. The results of
this simulation are rather interesting, as its dynamical evolution is similar 
to that calculated for the uniform density cloud models, that is, a 
cloud with $\rho(r)=\rho_0$ for all $r$, see for 
instance~\citet{NuestroApJ,NuestroRMAA}.
\section{Physical characterization}
\label{sec:resul}

We will discuss the behavior of some of the most important dynamical
variables related with the cloud collapse in the forthcoming
subsections. When necessary, we will restrict ourselves to consider only
the initial and final snapshots available in each simulation, as an
approximation to a complete time evolution of a dynamical variable.
It should be noted that the simulations were evolved by \citet{NuestroAA}
a little longer than we do for this paper.

As we will notice in Section~\ref{subs:dynevol}, in order
to study the results of the simulations, it
is enough to make iso-density plots for a slice of
particles around the cloud's equatorial plane. We present in
Figs.~\ref{Mos4A0}, \ref{Mos4A1}, \ref{Mos4A2} and \ref{Mos4A3} the main
results of the models to show the marked differences for each model. A more 
complete sets of results has already been presented by \citet{NuestroAA}.


\subsection{The cloud flattening}
\label{subs:dynevol}

The Plummer density profile assembles a very peculiar mass 
distribution; as it is pulled down
by the force of gravity towards the cloud center against the combined effect of
rotation and pressure, see Fig.~\ref{massprofile}.

The centripetal acceleration is given by $a_c=R_{\perp} \Omega$ for the spherical
cloud in rigid rotation with respect to the $Z-axis$, where $R_{\perp}$ is the
shortest distance from the particle to the rotation 
axis. As $R_{\perp}=R_0*\sin(\theta)$,
where $\theta$ is the polar spherical angle, then we have that $a_c$ has its
maximum value at the equator (where $\theta=\pi/2$) and
its minimum value at the poles (where $\theta=0$).

Every particle feels a centripetal acceleration, at least in the local reference
frame located on the particle, as a radial force, 
always opposing the radially attractive gravitational force. Thus, due to
the fact that this centrifugal force along the equator of the cloud is
greater than at the poles, the cloud contraction is faster at the poles
than at the equator; then the cloud evolves through a sequence of flatter
configurations parallel to the cloud equator and perpendicular to the
rotation axis. Numerical
simulations performed so far have proved that a uniformly rotating
molecular cloud, similar to the one considered here, contracts
itself in its innermost region during the isothermal regime to an almost
flat configuration, see for instance \citet{bb93,bb96} and \citet{review01}.
\subsection{Mass and density.}
\label{subs:density}

\begin{figure}
\begin{center}
\begin{tabular}{cc}
\includegraphics[width=3.0in]{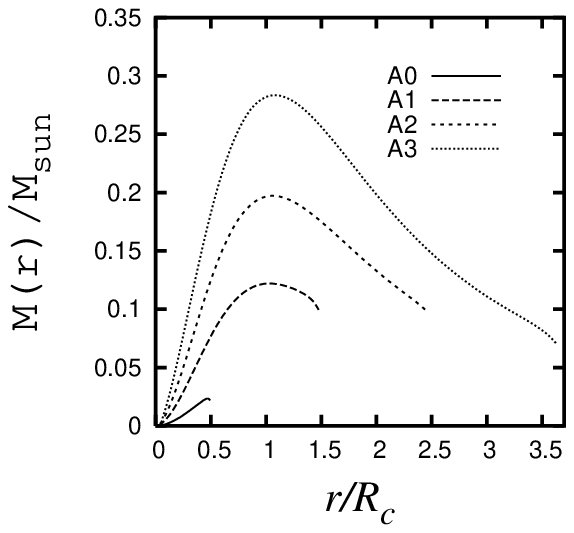}&\includegraphics[width=3.0in]{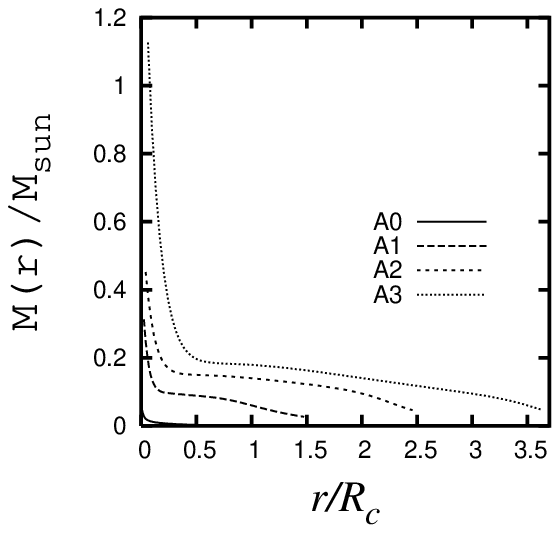}
\end{tabular}
\caption{The radial profile of the mass as a function of the
cloud's radius for all models for the initial snapshot (left)
and for the last snapshot available (right).}
\label{massprofile}
\end{center}
\end{figure}

According to the left panel of Fig.~\ref{massprofile}, initially more mass is
accumulated within the core boundary than in the gas envelope. As expected,
after most of the collapse has taken place, most of the mass had already
accumulated in the cloud center, as it can be seen in the right panel of
Fig.~\ref{massprofile}, where we show the mass radial profile for the last snapshot.

There is a characteristic time scale for the cloud collapse, which is given by

\begin{equation}
t_{ff} \approx \sqrt{\frac{3\, \pi} {32 \, G \, \rho}}\; .
\label{tffc}
\end{equation}
\noindent The free fall scale time $t_{ff}$ is defined by a characteristic
cloud density $\rho$. If we use the cloud average density $\rho_0$, 
the $t_{ff}$ will correspond to the time for a test particle falling freely from
the cloud surface to the cloud center. As our models have an increasing
radius, then the time we expect for the cloud to collapse ranges
from $8\, 744\; years$ for model $A0$; $45\, 436 \; years $ for
model $A1$; $ 97\, 764\; years $ for model $A2$ to $177\, 603\; years $ for
model $A3$. As we prefer to have only one scale time to normalize the
collapse history of all clouds, then we use the central core density, $\rho_c$,
which allows us to define the time $t_{ffc}= 38\, 460 \; years$.

There is clearly a first evolution stage, as it can be seen in
Fig.~\ref{evolucionmodelos}, in which the collapse proceeds very slowly until
the time has almost reached $t \approx t_{ffc}$. Shortly after, a stage of
a more rapid density increase in begins in which the peak density 
increases in a significant manner  until $\rho_{max}\approx \rho_c\, \times 10^{7}
\approx 3 \times 10^{-11} \, gr/cm^3$.

The model $A3$ collapse takes a longer time than the others because it has
more pressure support, more mass and more envelope extension than the other
models. Additionally, we note that a smaller number of particles in
model $A3$ achieves higher densities than in model $A0$, as illustrated in
the right panel of Figure~\ref{evolucionmodelos}, where a particle distribution
characterized by the peak density is shown for the last snapshot available in each
simulation. To interpret this plot, consider a vertical line, as the
one labeling the fraction $f=0.8$; then, this means that $80 \%$ of the particles in
model $A0$ have a density greater than
$\log_{10}\left( \rho_{max}/\rho_0\right)\approx 2 $, which translates
into $\rho_{max}=3.0 \times 10^{-16} \, gr / cm^3 $; whereas in  model $A3$ for the
same fraction of particles, $80 \% $ have a density higher
than $\rho_{max}= 4.75 \times 10^{-20} \, gr / cm^3 $, which is a very low
value indeed. We observe therefore that the model $A0$ collapses faster than the
model $A3$.

\subsection{Acceleration.}
\label{subs:Acceleration}

Let us consider now the acceleration generated by the particular mass
distributions assembled in the Plummer
clouds. We show the total acceleration radial component
evolution as a function of radius in Fig.~\ref{aceletotal}. These accelerations have
been calculated by dividing the cloud into a fixed number (30) of spherical shells
and averaging the radial accelerations of all the particles
contained in the same shell.  As it can be seen in the left panel of
Fig.~\ref{aceletotal}, the curve has a local minimum, indicating that
there is a shell of material which feels the highest gravitational
attraction in the cloud.

If $r_{min}$ is the radius of the shell with the
highest gravitational pull, then for those shells farther away, that is
with $r > r_{min}$, their total acceleration begins to increase making the
gravitational force acting on these layers outside the core not
very relevant. It was analitically demostrated by \citet{bh04} that 
the radial acceleration for $2d$ disks diverges at $r=R_0$, but that  
for a finite thickness disk, the divergence would not occur.  

As expected, the behavior of accelerations show significant differences in
the final stage of the collapse process, as it can be appreciated in the right
panel of Fig.~\ref{aceletotal}. The hydrodynamic pressure is clearly dominant in
the cloud center, where it even shows a clear tendency towards gas
expansion. As a consequence of the cloud rigid body rotation, a term of
centripetal acceleration -$\Omega^2 \, r$ appears, directed toward
the cloud center, which makes it very difficult to increase the magnitude 
of the total acceleration.

\subsection{Angular momentum and $aem$ ratio.}
\label{subs:momentum}

As there is no external force acting upon the cloud, the total angular momentum must
be conserved. We verify this conservation property by summing up all the
$SPH$ particles of a run: $\vec{J}=\Sigma_{i=1}^{N}\, m_i\, \vec{r}_i \times \vec{v}_i$,
and using all the snapshots obtained in each simulations, as it can
be seen in both panels of Fig.~\ref{ConserAngMom}. We see in this plot that the
cloud for model $A0$ has the smallest specific angular momentum, while
its $aem$ ratio is the highest. These values can be easily
explained for model $A0$, because it has
less mass and its size is smaller than in the other
models. Nevertheless, the $aem$ ratio value
is almost the same for the rest of the models.

\subsubsection{Radial profile.}
\label{subsubs:radialprofile}

The angular momentum for a rigid body is given by the $J=I \ \Omega$ relation,
where $I$ is the moment of inertia and $\Omega$ is the angular velocity.
The cloud in our models is a rigid sphere-like in the initial snapshot; we will now try
to generalize this simple mathematical relation to our cloud models for later
times. We would have in such case

\begin{equation}
\log_{10} \left( J/M \right) =\zeta \,\log_{10} \left( r/R_0 \right) + \log_{10} \left( \Omega \right)
\label{eq:perfilmomento}
\end{equation}
\noindent with $\zeta$ being a constant. We have also calculated both the angular
momentum and the $aem$ ratio radial profile, aiming to figure out to what extend
this relation remains valid in the cloud evolution. As was done for the
acceleration calculation, this task was carried out by dividing the cloud in thin
shells, to add the contribution of each particle within the shell afterwards, so
that at the end of the task we end up with the momentum and
mass for every shell. We can see, by
applying this procedure to the first snapshot of each
simulation, that the relation \ref{eq:perfilmomento} is initially well justified, as
it is shown in the left panel of Fig.~\ref{JenMPerfil}.

Now, in the right panel of Fig.~\ref{JenMPerfil} we present the results
of the same calculations on shells, but now for the last snapshot available in
each simulation. As expected, at the final time of evolution we observe
that some kind of differential rotation regime is present, above all,
for the cloud outermost regions. At that point, relation \ref{eq:perfilmomento} 
is no longer valid, as the cloud geometry and the mass distributions have already
changed. Indeed, the cloud moment of inertia has changed somewhat due to the
process of material accumulation at the cloud center. The effects of this accretion
process look more dramatic for the case of the $aem$ ratio radial profile, as it
can be seen in Fig.~\ref{aemPerfil}. The reason for this behavior is again the
mass accumulated in the cloud central region, as the $aem$ ratio magnitude within
a gas radial shell goes as the squared of the mass; then its magnitude is
significantly reduced.

\subsubsection{Correlation with the particle peak density.}
\label{subsubs:correpartpeakdens}

We change the independent variable in the
preceding Figures from cloud radius to density. Let us start by looking at
Fig.~\ref{DistLyRho}, where we show the specific
angular momentum and $aem$ ratio distribution against the particle density, in
the last snapshot available for each of the different runs. It is clearly seen
that as the $SPH$ particle eventually acquires a greater density, its momentum
and $aem$ ratio decrease.

Regardless of the peak density value, the particles are giving up part of
their angular momentum as a consequence of both the shear viscosity presence and
the decreasing value of their radial distance $\vec{r}$ to the coordinates origin
and, because in some cases, this origin coincides with the center of the cloud densest
central region.

Additionally, it is interesting to note that there is a more pronounced drop of
the $aem$ ratio in the model $A3$ than for the model $A0$, as it can be seen in
the right panel of Fig.~\ref{DistLyRho}. Even though the collapse is faster
for the model $A0$ than for the model $A3$. Shear effects are probably less
important for model $A0$ than for model $A3$, as its
mass and velocities observed in the cloud central region are
smaller for model $A0$ than for model $A3$, as we will see below.

However, we observe in Fig.~\ref{DistLyRho} that there is a final stage in which
the loss of the specific angular momentum and $aem$ ratio is less severe for all the
models; furthermore, for the model $A3$ one can see even a trend toward a
recovery in the value for the last part of the curves of angular momentum and
$aem$ ratio.  This behavior can be explained because the envelope of the model $A3$
increases, and its particles are more distant (a larger $r$) from the center, so
their angular momentum must still be higher. There is also another reason,
which is due to the appearance of the fragments orbital motion, as will be
discussed in Section~\ref{subs:defclumpsandfrag}.

We show the velocity field distribution of all those particles
located within the cloud central region in Fig.~\ref{VelDistri}, in order to
shed more light into the marked differences obtained at the cloud center,
according to the cloud model and at the time reached by the last available snapshot.
\subsubsection{Rate of change of the angular momentum with the cloud radius.}
\label{subsubs:ratechanwithcloudrad}

Let us end this section by considering the change in the angular momentum radial
profile for a particle located initially at $\vec{r}+\vec{ \Delta r}$ and moving
towards an innermost radial shell $\vec{r}$. The new angular momentum is

\begin{equation}
\vec{L} ( \vec{r}+\Delta \vec{r}  ) = m \, ( \vec{r}+\Delta \vec{r} ) \, \times \, ( \vec{v}+\Delta \vec{v} )
\label{defmomentoporpart}
\end{equation}
\noindent  where $m$ is the mass of the particle. Then, replacing the kinematic
relations $\Delta \vec{r}= \vec{v} \, \Delta t$ and
$\Delta \vec{v}= \vec{a} \, \Delta t$ into Eq.~\ref{defmomentoporpart}, we obtain
the following differential equations which are valid only to first order,

\begin{equation}
\frac{d \, \vec{L} ( \vec{r} )} { d \; r}  = \frac{m}{\dot{r}}  \, \vec{r}\, \times \vec{a}
\label{eqmomentoenr}
\end{equation}
\noindent  where $\dot{r}$ is defined as $\frac{d \, r}{ d\, t}$;
this function obviously depends on the very particular way in which
the gas particles accretion is taking place. Furthermore, the Eq.\ref{eqmomentoenr}
clearly indicates that there would be no change in the
angular momentum with respect to $r$ for models with a purely radial
acceleration (as a cloud having all its particles moving in a homogeneous circular
motion ). There would indeed be a change in the angular momentum only for those
particles having a non-zero tangential (centrifugal) acceleration. This would be the
case if either shear viscosity is present or if a redistribution of forces occurs in
the cloud central region as a consequence of a change in the clump geometry, as we
discuss in Section~\ref{sec:dis}.

It is beyond the limited scope of this paper to consider the equation
describing the change of the angular momentum for those particles falling into the
cloud center, which tentatively is still an unknown and perhaps very complicated
issue. However, we can take advantage of our simulations for measuring the way
in which those particles being accreted are losing their angular momentum. We
have selected a set of particles for this purpose, which have already reached the
cloud innermost region at the last available snapshot for each model. Subsequently,
we followed this set of particles -in as many previous snapshots as
possible- along their path into the densest central gas clump. As an
instance, in Fig.~\ref{SeguiDer} we show the rain of particles falling off
into the formed clumps. It is interesting to note that there is a very marked
fall in the angular momentum value only when the particles are really
close to the densest clump, as it can be appreciated in
Fig.\ref{SeguiDerLvsR}. It is therefore the particles of the innermost
disk which are the most relevant for the momentum interchange.

We clarify that in Fig.\ref{SeguiDerLvsR} a dot corresponds to a $SPH$ particle of
the simulation; thereby, the shaded region in these plots indicates an important
accumulation of particles. 

\subsection{Energy ratios.}
\label{subsec:dynprop}

As we previously mentioned in Section\ref{subs:density}, our collapse models stop
evolving when the first formed matter aggregates reach a
peak density around $10^{-11} \, gr/cm^3$. Those gas aggregates
can be identified as protostellar cores, already. Nowadays, it is well established
that these protostellar cores physical characteristics are more likely to be
inherited by the stars that might result from them if they could collapse further
until peak densities around $10^{-1} \, gr/cm^3$ are reached. It is therefore very
important to study those protostellar aggregates physical properties, as we do below.
\subsubsection{Defining fragments.}
\label{subs:defclumpsandfrag}

We define the center, $\vec{x}_{center}$, of a matter aggregate as the particle
with the highest density in the region where the aggregate is located. We then find
all the particles, let us say $N_s$, which have a density above (or equal to)
some minimum density value $\rho_{min}$ and which, at the same time, are located within
a given maximum radius $r_{max}$ from the aggregate center. We can define a region
of matter with this set of $N_s$ particles from which we can estimate the integral
properties, as it will be escribed in Section\ref{subs:calproc}. When the
two cutting parameters, $\rho_{min}$ and $r_{max}$, are taken into
account at the same time for selecting particles, then the aggregate of matter will
be referred as a {\it fragment}. We plot the centers of these matter
aggregates in Fig.~\ref{CentrosForModels}, for each simulation.
\subsubsection{Calculation procedure.}
\label{subs:calproc}

We now show the way in which we can estimate the energy ratios $\alpha$ and $\beta$
for a set of $N_s$ particles defining a fragment. The first step is to obtain the
density and the gravitational potential for every particle $i$ due to the presence
of all others particles $j\neq i$.

We use the smoothing kernel for calculating the particle density
$i$ by means of $\rho_i \equiv \rho(\vec{r_i})=m\,
W_1(\vec{r}_i,h)$, where $W_1(\vec{r}_i,h)$ is the spline kernel
given in Eq. A.1 of \citet{gadget1}. We use another kernel for the gravitational
potential,  such as $\Phi_i\equiv \Phi(\vec{r}_i)=G\,
\frac{m}{h}\, W_2(\frac{\vec{r}_i}{h})$, where
the kernel $W_2$ is now given in Eq. A.3 of the same reference.
The softening length $h$ appearing in these two kernels sets the
neighborhood on the point $\vec{r}$, outside of which no particle can
exert influence on $\vec{r}$; that is, for $r>h$ both kernels
vanish: $W_1\equiv 0$ and $W_2\equiv 0$. We use several values for
$h$, looking forward to have a number of neighbor particles for any
point (or particle) greater than or equal to $50$.

We approximate the thermal energy of the clump by calculating the sum over all
the $N_s$ particles, that is

\begin{equation}
E_{therm}=\sum_{i=1}^{N_s} \, \frac{3}{2} \,
\frac{P_i(\rho)\,m_i}{\rho_i} \, ,
\end{equation}
where $P_i$ is the pressure associated with particle $i$ with density
$\rho_i$ by means of the equation of state given in Eq. \ref{beos}. Similarly, the
approximate potential energy is

\begin{equation}
E_{pot}= \sum_{i=1}^{N_s}\, \frac{1}{2}\, m_i \, \Phi_i \, .
\end{equation}

Although a bit more complicated, the rotational energy of $N_s$ particles can
be calculated as follows with respect to the $Z-axis$ of the located
clump. Let $\vec{x}_i$ and $\vec{v}_i$ be the position and velocity of particle $i$
in the gadget2 coordinates. Thereby the coordinates of those particles in the clump
with respect to the clump center are $\vec{u}_i=\vec{x}_i-\vec{x}_{center}$. The
azimuthal angle $\phi_i$ associated with the particles rotation with respect to
the $Z-axis$ can be calculated by taking the ratio of particle coordinates
projection with the unitary vectors
$\hat{\i}=(1,0,0)$ and $\hat{\j}=(0,1,0)$, that is $\phi_i=\arctan
\left( {\vec{u}_i\cdot \hat{\j}}/{\vec{u}_i\cdot \hat{\i} }\right) $. 
The rotational energy can be thus estimated by taking the
projection of the velocity along the unitary azimuthal vector
$\hat{e}_{\phi_i}=-\sin(\phi) \hat{\i} + \cos(\phi) \hat{\j}$, that
is

\begin{equation}
E_{rot}=\sum_{i=1}^{N_s}\, \frac{1}{2}\, m_i
\, \left( \vec{v}_i\cdot \hat{e}_{\phi_i} \right)^2 \,.
\end{equation}
\subsubsection{Calculated properties for fragments.}
\label{subs:propclumpsandfragments}

We arbitrarily chose the cutting density $\rho_{min}=1.40 \, \times 10^{-17} \; gr/cm^3$
for defining a clump, which corresponds to $100$ times the cloud average
density for model $A3$, see Table~\ref{tab:modelos}. Any clump with this
cutting density will in general include about $1.5\% $ of the total number
of particles in the simulation.

We sum to the mass and to the forming clump angular momentum, going forward
through as many snapshots as possible in each simulation, the contribution of
all those particles having a density higher than $\rho_{min}$.  As a matter of fact,
we show in Fig.~\ref{AngMomVSMasalrhomin} how the angular momentum 
and the mass of the clump
evolve as more particles enter into the forming clump. We have
added a $C$ in these plots to emphasize that we are not only using the
second cutting parameter $r_{max}$.  The left panel shows the specific angular
momentum, while the right panel shows the $aem$ ratio. We see that
very few particles in model $A0$ reach densities higher than $\rho_{min}$ long
before other particles; whereas in model $A3$, the collapse is more uniform, in
such manner that many more particles come to be part of the forming clump at
the same time.

Another observation from these two panels of Fig.~\ref{AngMomVSMasalrhomin} is
that as the gas envelope extension increases, the specific angular momentum
increases as well, but the $aem$ ratio decreases.  This is because
the first particles joining the forming clump bring more angular momentum than
those particles that collapse afterwards, which provide more mass to the clump 
than angular momentum. As the $aem$ ratio is more sensitive to the mass
contained in the clump, the $aem$ ratio magnitude falls as the new clump
is forming.

Moreover, Fig.~\ref{AngMomVSMasalrhomin} indicates that in the model $A3$
more particles are still located in the envelope, and most of them are still keeping
a large angular momentum, as it can be
seen in Fig.~\ref{DistLyRho}; meanwhile, a higher proportion of particles have
already entered the phase of most advanced collapse in the model $A0$ and, as a
consequence, have already lost most of their angular momentum.

We report the energy ratios calculated by using the two cutting parameters and by
the application of the calculation procedure outlined in
Section\ref{subs:defclumpsandfrag} to the last
snapshot obtained for each simulation in Table~\ref{tab:PropIntFrag}. The entries
of Table~\ref{tab:PropIntFrag} are as follows. We show the model label for
which we are going to account for only the formed central clump in column 1. We
indicate the number of particles ($N_s$) entering into the set of particles used
for approximating the energy ratios calculation in column 2. We show in
columns 3 and 4 the energy ratios $\alpha_f$ and $\beta_f$ as previously defined
in Section~\ref{subs:calproc}. We emphasize that the minimum density values
expressed in terms of the $\log_{10}\left( \rho_{min}/\rho_0\right) $ have been
taken to be $4.0$ for all the models; this is the lowest value that a particle entering
in the set can indeed have. Moreover, we indicate the maximum radius expressed in
terms of the size of the simulation box, $2\, R_0$, which was fixed to $0.01$ in
order to delimitate the clump radial extension.

\begin{table}[ph]
\caption{Energy ratios for the central fragment. }
{\begin{tabular}{|l|c|c|c|c|c|c|}
\hline
{Model} & $N_p$ &  $|\alpha_f |$ & $|\beta_f |$ & $|\alpha_f |$ + $|\beta_f |$ \\
\hline
  A0    &  689951 & 0.22 & 0.18 & 0.4 \\
\hline
  A1    &  248831 & 0.25 & 0.20 & 0.45 \\
\hline
 A2     &  109924 & 0.27 & 0.13 & 0.4 \\
\hline
  A3    &  119326 & 0.25 & 0.24 & 0.49 \\
\hline
\end{tabular}
\label{tab:PropIntFrag}}
\end{table}

\subsubsection{Global Properties.}
\label{subs:GlobalProp}

As we have already applied the procedure outlined in Section ~\ref{subs:calproc}
in previous publications, it is now clear that the numerical results for the energy
ratios unfortunately depend on the values chosen for the two cutting parameters,
$\rho_{min}$ and $r_{max}$, as we inevitably commit certain ambiguity in defining
the clump boundaries. Furthermore, with this procedure we obtained information only
about the physical state of each clump or fragment, separately.

We calculated the energy ratios in this section using all the particles in
each simulation, to avoid cutting ambiguities. 
We follow again as inthe previous Section~\ref{subs:calproc}, but in 
this case, we have $N_s=N$. We have
added the subscript $w$ on the plot, to distinguish those quantities
calculated when using all the simulation particles.

It is important to emphasize that there is a clear advantage in taking all the
particles for this calculation, as the new clumps formation resulting from
the gravitational collapse can be recorded in the behavior of the $\alpha$
and $\beta$ variables, as we will describe in this Section.

When a clump begins to form, the pressure of their constituent particles 
increases, and we would expect that the $\alpha$ variable would also show an increase
in spite of the fact that the gravitational potential is also 
growing in magnitude. Since now we
do not care where the particles are located in the cloud for the purpose of
measuring the $\alpha$
and $\beta$ values, we can capture the formation of new clumps wherever they
start. We therefore have a tool for recording an imprint left by the
occurrence of fragmentation in the cloud on these dynamical variables.

Let us take a look at Fig.~\ref{IntPropSnap}, where
the curves behavior $\alpha$ {\it vs} $\beta$ are
somewhat different for each simulation, indicating
the different outcomes of each cloud model. However, in these
panels there are clearly common features
in all the curves, establishing a very strong similarity between
the pairs of models $A0$ with $A1$ and $A2$ with $A3$. This association in
pairs of models is obviously a consequence of the gas envelope extension
on the simulation outcome.

\begin{figure}
\begin{center}
\centerline{\epsfig{width=150mm, figure=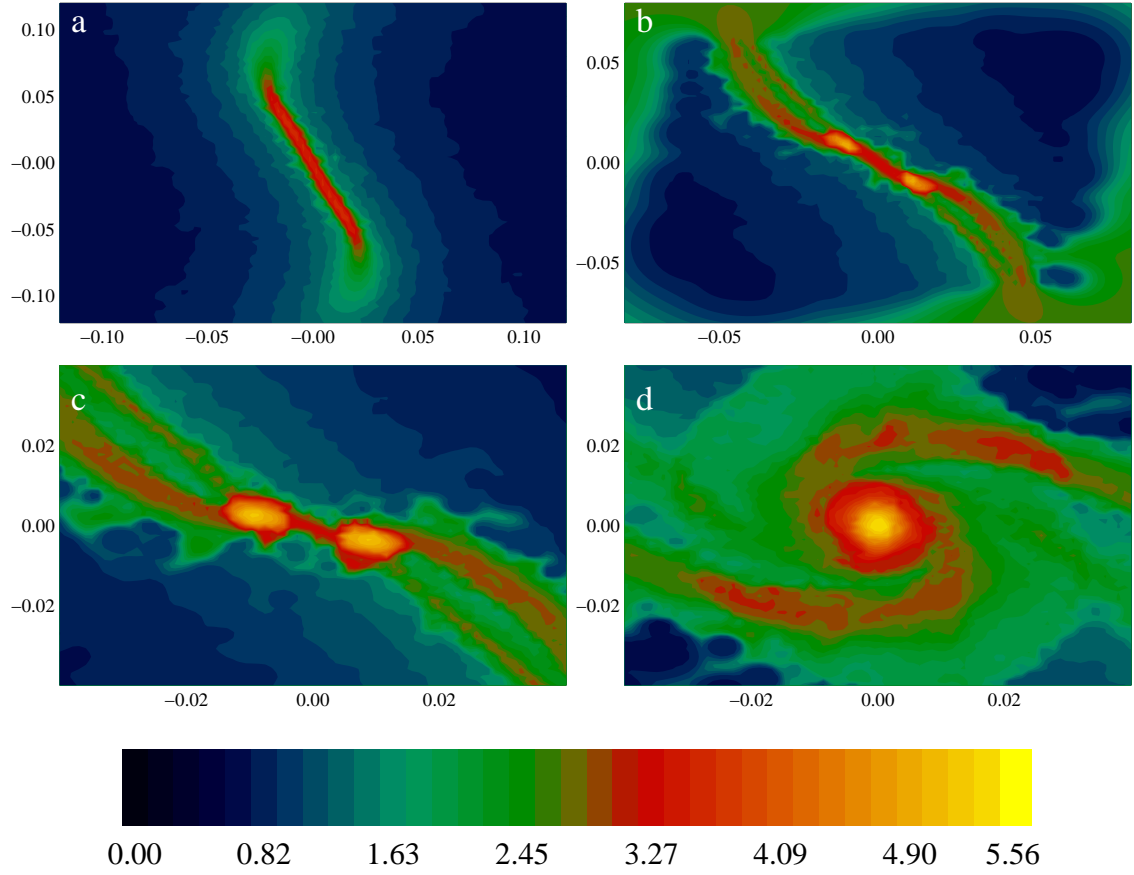}}
\caption{Isodensity curves of the cloud's mid-plane for model $A0$ when
the distribution of particles reaches a peak density of
(a) $\rho_{max}= 3.3 \, \times 10^{-13}\, gr/cm^3$ at time $t=1.64 \, \times 10^{12} \, sec$
(b) $\rho_{max}= 5.8 \, \times 10^{-12}\, gr/cm^3$ at time $t=1.66 \, \times 10^{12} \, sec $
(c) $\rho_{max}= 1.3 \, \times 10^{-11}\, gr/cm^3$ at time $t=1.68 \, \times 10^{12} \, sec $
(d) $\rho_{max}= 2.2 \, \times 10^{-11}\, gr/cm^3$ at time $t=1.71 \, \times 10^{12} \, sec$. }
\label{Mos4A0}
\end{center}
\end{figure}

There is a first stage marked by the labels 1 and 2, which points to the fact
that the early collapse evolution of all the cloud models proceeds in an identical
manner; in this first stage the $\alpha$ is decreasing as a consequence of the
systematic increase of the gravitational potential for the cloud
central  particles. The 1-2 stage takes about a free fall time for each cloud,
at a time at which the cloud central part has already lost its initial spherical
symmetry, because the densest particles have found a place in a narrow slice
of matter ( the filament ) around the equatorial plane, occupying approximately
up to 10 \% of the original cloud size.  The upper left panel of Fig.~\ref{Mos4A0}
corresponds to the end of this first stage 1-2, as it is indicated in the
panel labeled with $A0$ of Fig.~\ref{IntPropSnap}.

\begin{figure}
\begin{center}
\centerline{\epsfig{width=150mm,figure=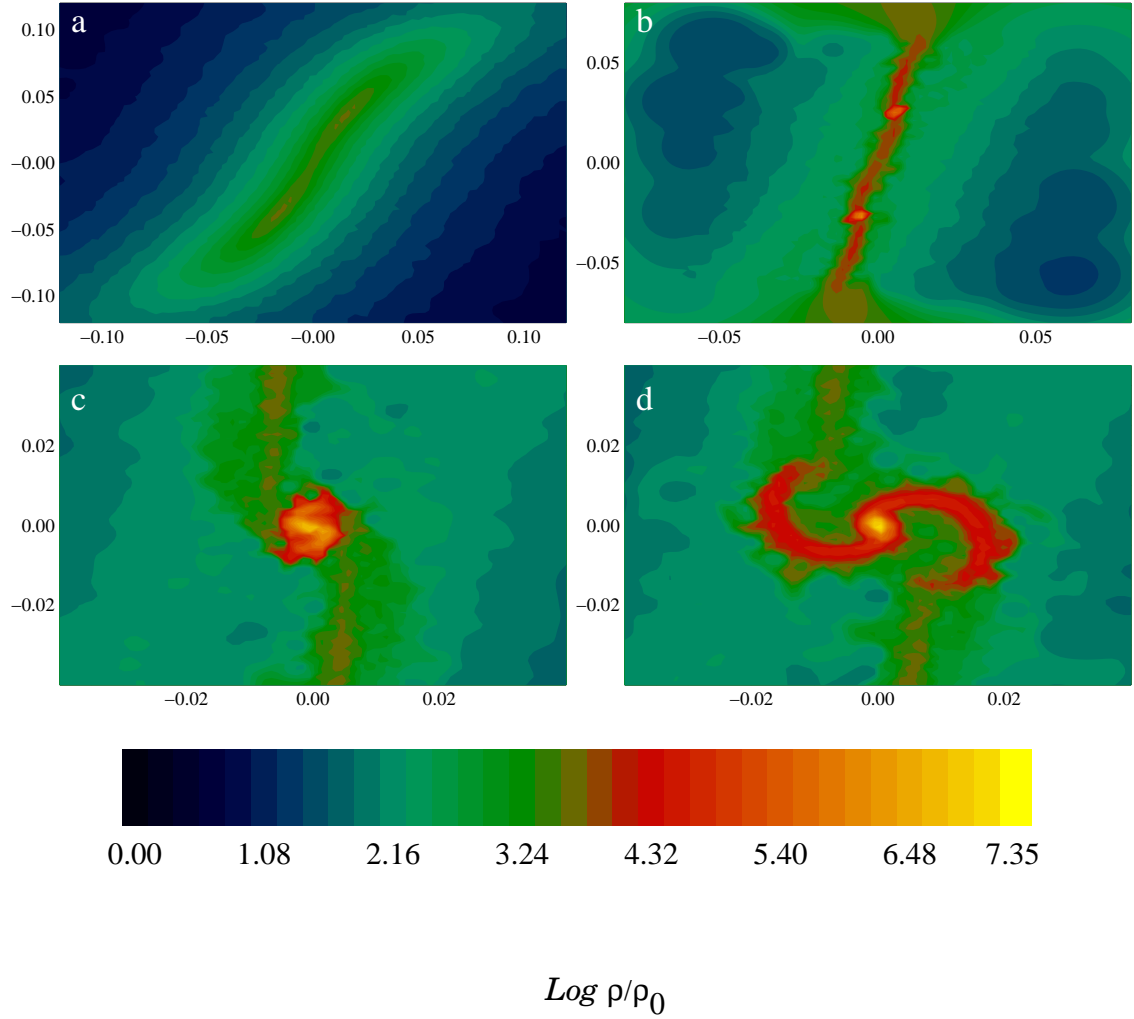}}
\caption{The same as Fig.~\ref{Mos4A0} but for $A1$ when
(a) $\rho_{max}=1.15  \, \times 10^{-14}\, gr/cm^3$ at time $t=1.78 \, \times 10^{12} \, sec$
(b) $\rho_{max}=1.29  \, \times 10^{-11}\, gr/cm^3$ at time $t=1.88 \, \times 10^{12} \, sec $
(c) $\rho_{max}=3.15  \, \times 10^{-11}\, gr/cm^3$ at time $t=1.95 \, \times 10^{12} \, sec $
(d) $\rho_{max}=5.29  \, \times 10^{-11}\, gr/cm^3$ at time $t=1.96 \, \times 10^{12} \, sec$. }
\label{Mos4A1}
\end{center}
\end{figure}

\begin{figure}
\begin{center}
\centerline{\epsfig{width=150mm,figure=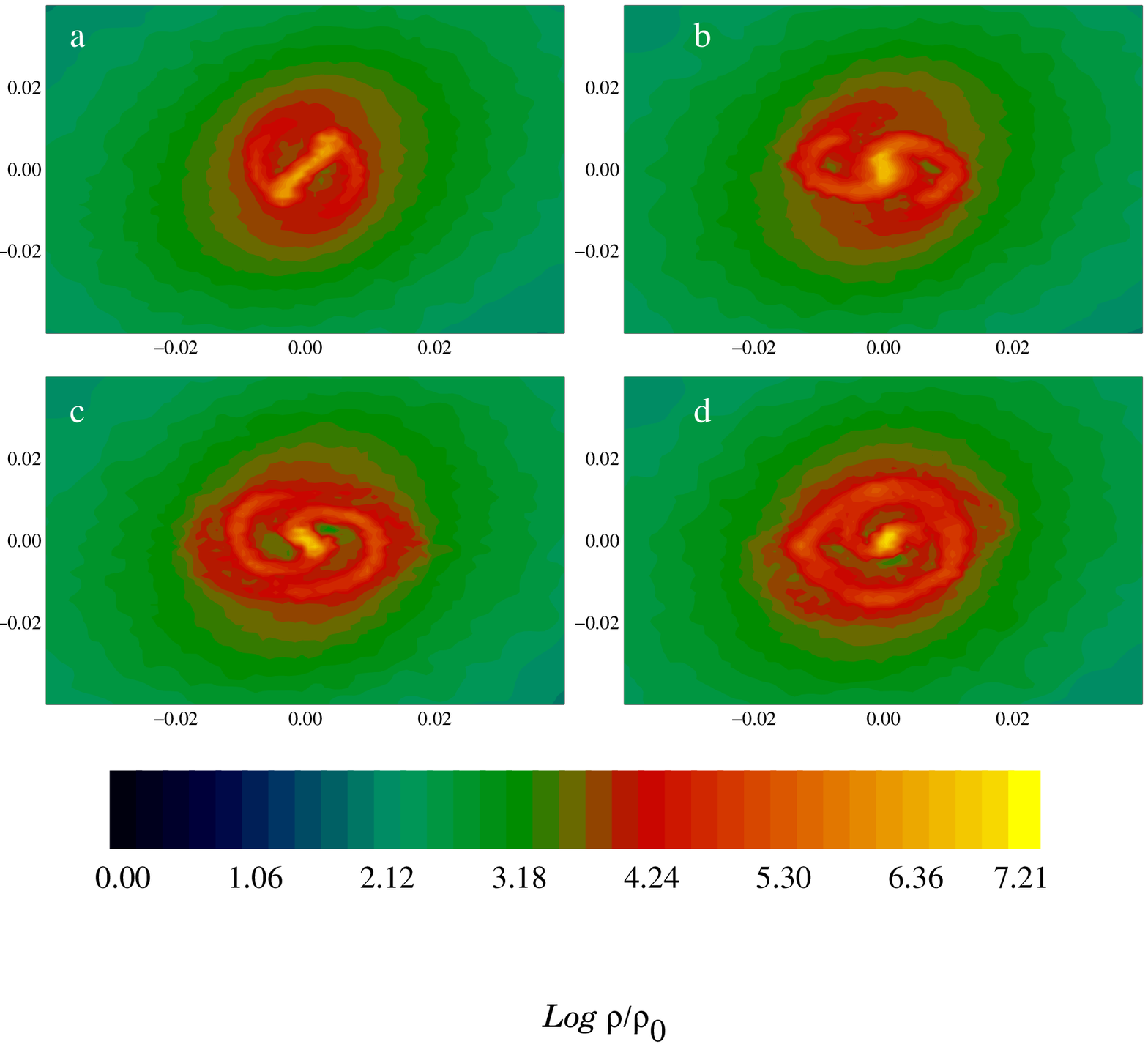}}
\caption{The same as Fig.~\ref{Mos4A0} but for $A2$ when
(a) $\rho_{max}=1.8  \, \times 10^{-12}\, gr/cm^3$ at time $t=1.82 \, \times 10^{12} \, sec$
(b) $\rho_{max}=4.4  \, \times 10^{-12}\, gr/cm^3$ at time $t=1.85 \, \times 10^{12} \, sec $
(c) $\rho_{max}=7.1  \, \times 10^{-12}\, gr/cm^3$ at time $t=1.87 \, \times 10^{12} \, sec $
(d) $\rho_{max}=8.6  \, \times 10^{-12}\, gr/cm^3$ at time $t=1.89 \, \times 10^{12} \, sec$. }
\label{Mos4A2}
\end{center}
\end{figure}

\begin{figure}
\begin{center}
\centerline{\epsfig{width=150mm,figure=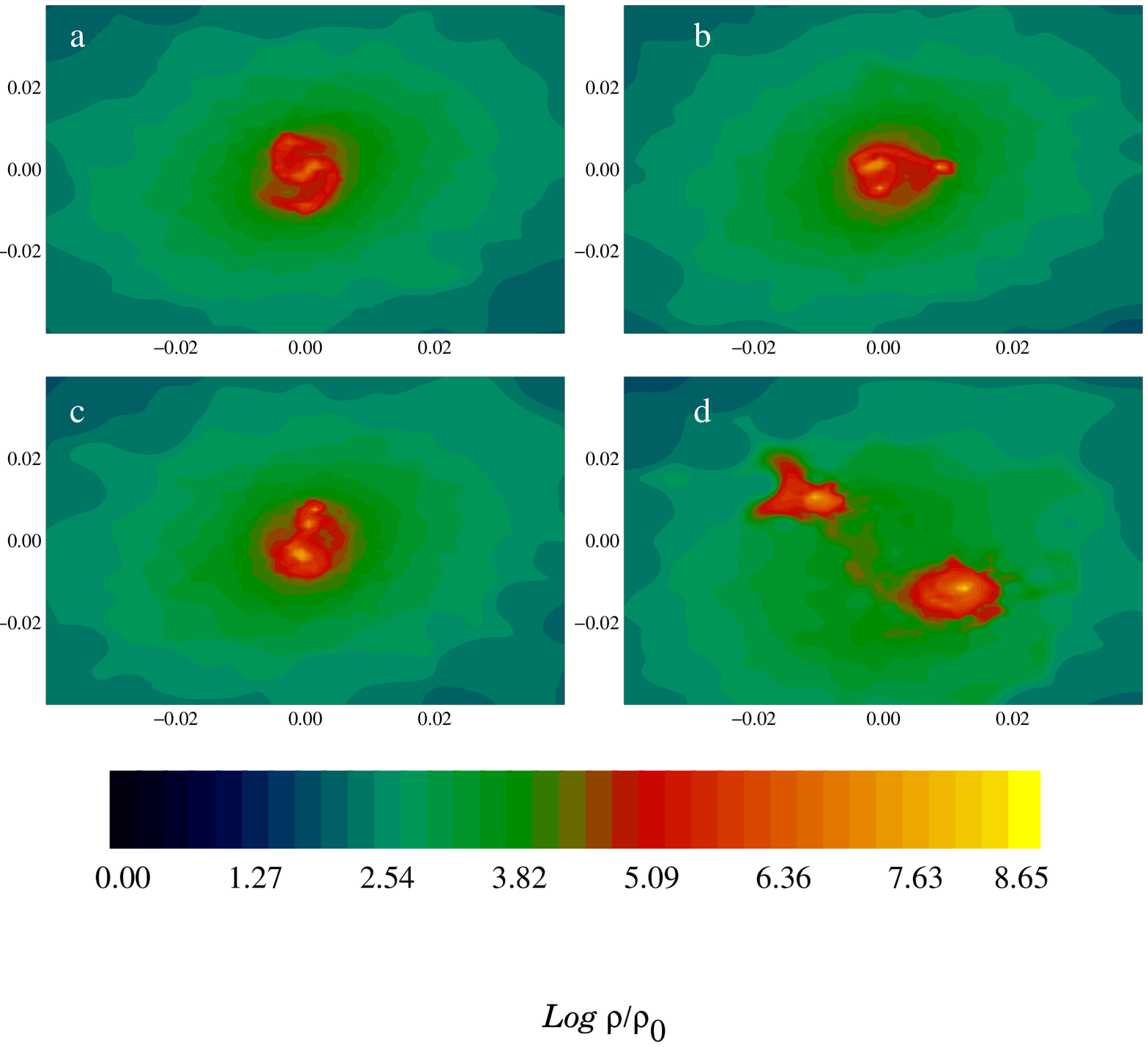}}
\caption{The same as Fig.~\ref{Mos4A0} but for $A3$ when
(a) $\rho_{max}=7.74  \, \times 10^{-12}\, gr/cm^3$ at time $t=1.82 \, \times 10^{12} \, sec$
(b) $\rho_{max}=1.22  \, \times 10^{-11}\, gr/cm^3$ at time $t=1.86 \, \times 10^{12} \, sec $
(c) $\rho_{max}=1.94  \, \times 10^{-11}\, gr/cm^3$ at time $t=1.88 \, \times 10^{12} \, sec $
(d) $\rho_{max}=1.67  \, \times 10^{-10}\, gr/cm^3$ at time $t=2.19 \, \times 10^{12} \, sec$. }
\label{Mos4A3}
\end{center}
\end{figure}

\begin{figure}
\begin{center}
\begin{tabular}{cc}
\includegraphics[width=3.0in]{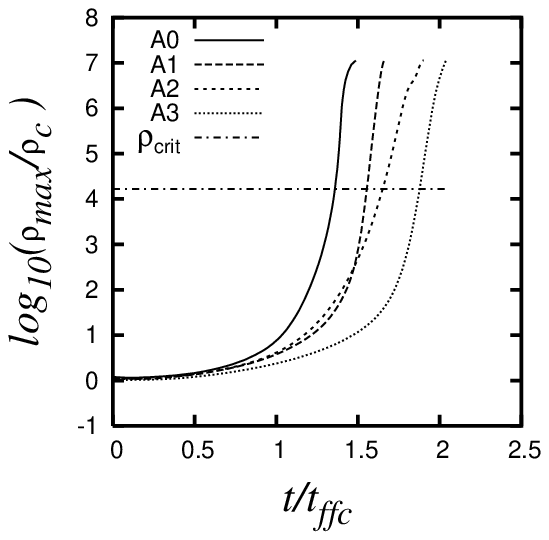}&\includegraphics[width=3.0in]{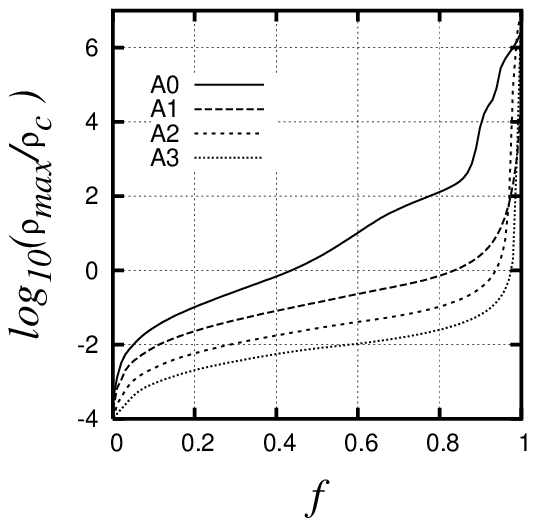}
\end{tabular}
\caption{Time evolution of the peak density of the cloud for all
models (left). For the last snapshot available in each simulation, we show the
fraction $f$ of $SPH$ particles with a peak density
higher than a given peak density as shown in the vertical axis (right). }
\label{evolucionmodelos}
\end{center}
\end{figure}

\begin{figure}
\begin{center}
\begin{tabular}{cc}
\includegraphics[width=3.0in]{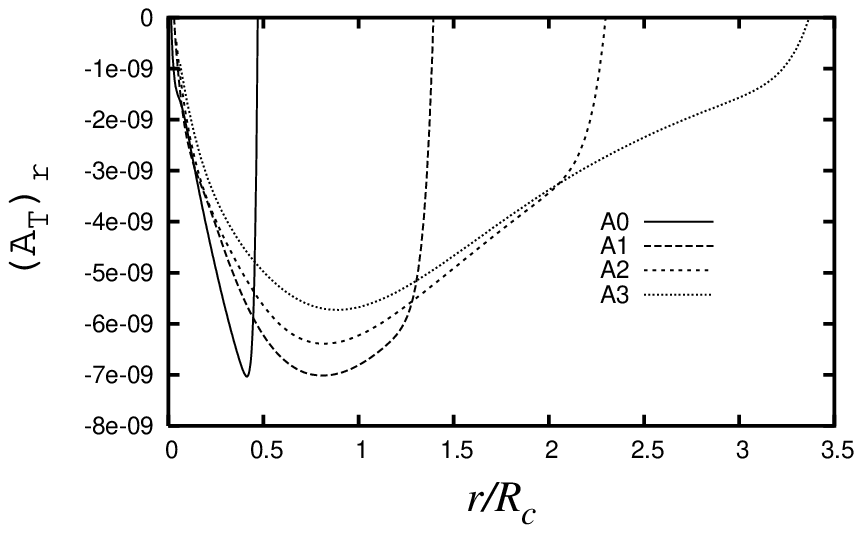}&\includegraphics[width=3.0in]{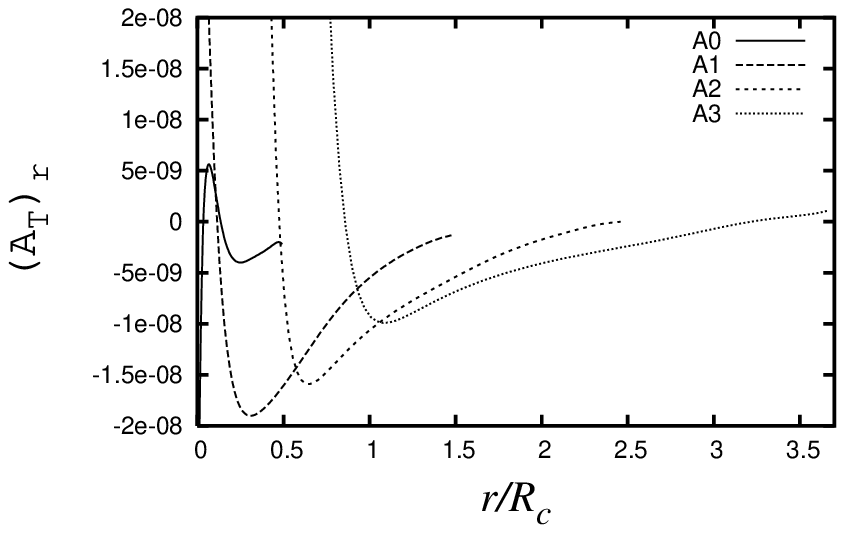}
\end{tabular}
\caption{The radial projection of the total acceleration as a function of the
cloud's radius for all models for the initial snapshot (left)
and for the last snapshot available (right).}
\label{aceletotal}
\end{center}
\end{figure}

\begin{figure}
\begin{center}
\begin{tabular}{cc}
\includegraphics[width=3.0in]{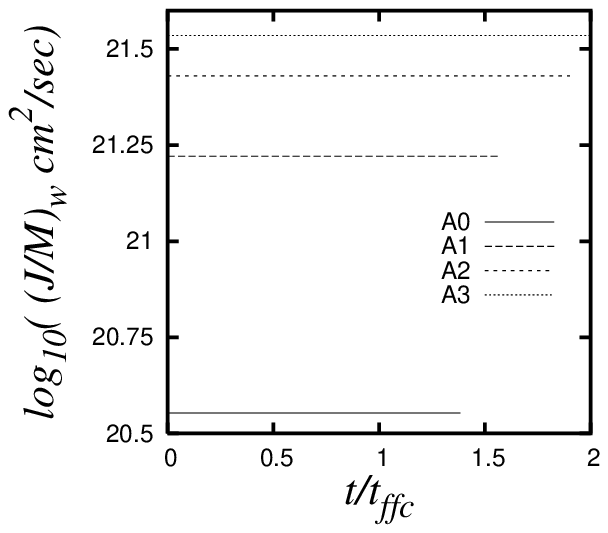}&\includegraphics[width=3.0in]{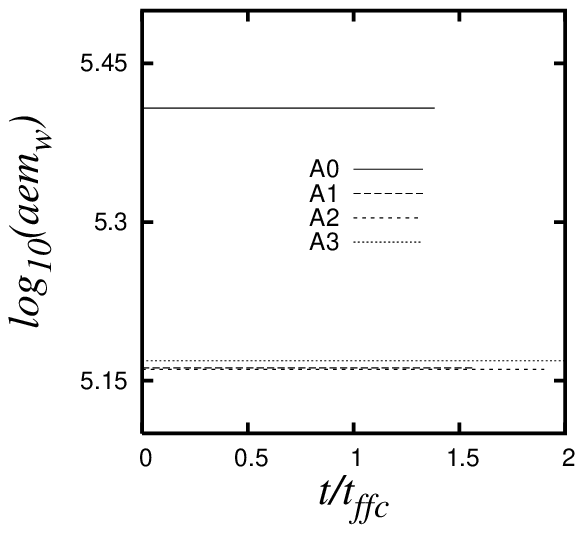}
\end{tabular}
\caption{
The time evolution of the specific angular momentum $J/M$
(left) and of the $aem$ ratio (right). Note the good level of
conservation of both of these quantities for all the collapse models.}
\label{ConserAngMom}
\end{center}
\end{figure}

\begin{figure}
\begin{center}
\begin{tabular}{cc}
\includegraphics[width=3.0in]{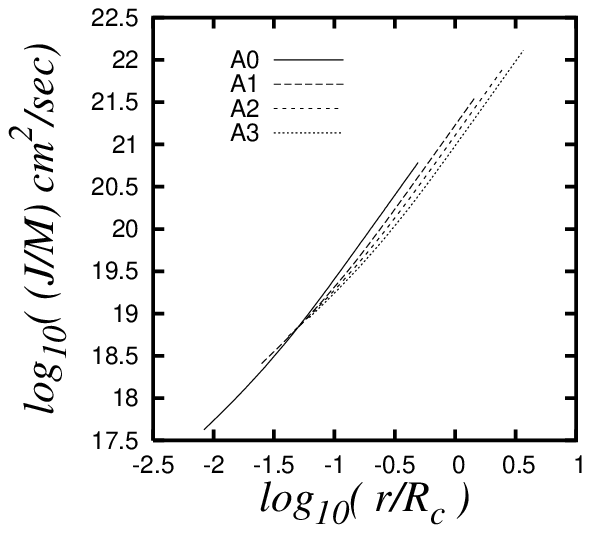}&\includegraphics[width=3.0in]{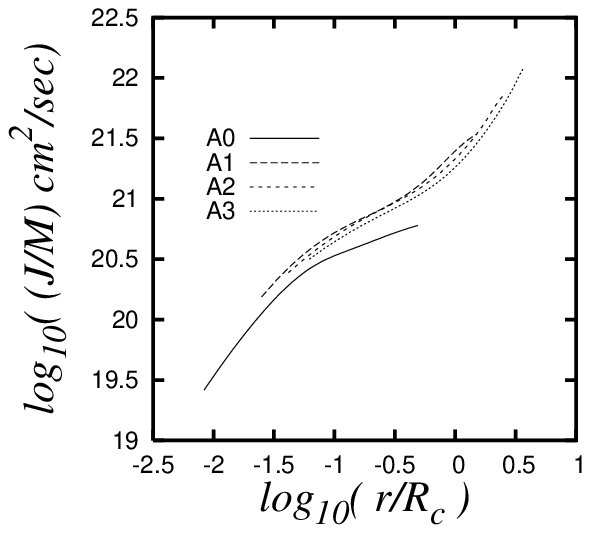}
\end{tabular}
\caption{Radial profile distribution of the specific angular momentum for
all collapse models, for the first snapshot (left) and for the last
snapshot available (right).}
\label{JenMPerfil}
\end{center}
\end{figure}

\begin{figure}
\begin{center}
\begin{tabular}{cc}
\includegraphics[width=3.0in]{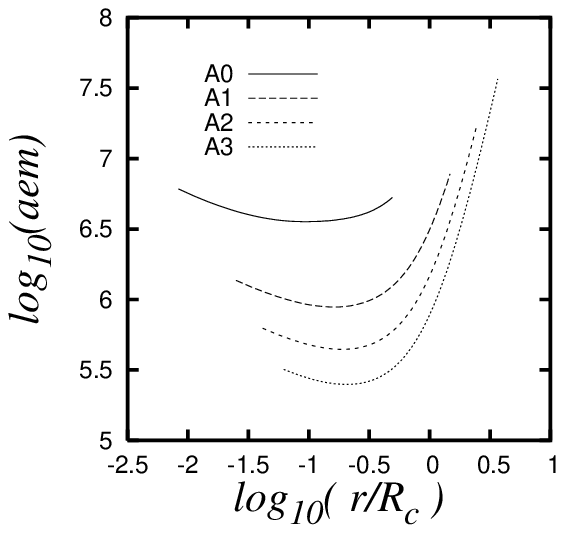}&\includegraphics[width=3.0in]{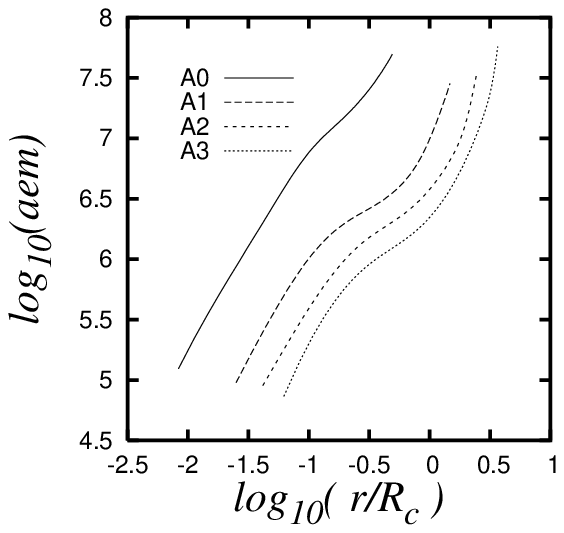}
\end{tabular}
\caption{ Radial profile distribution of the $aem$ ratio for the
initial snapshot (left) and  for the last snapshot available (right).}
\label{aemPerfil}
\end{center}
\end{figure}

\begin{figure}
\begin{center}
\begin{tabular}{cc}
\includegraphics[width=3.0in]{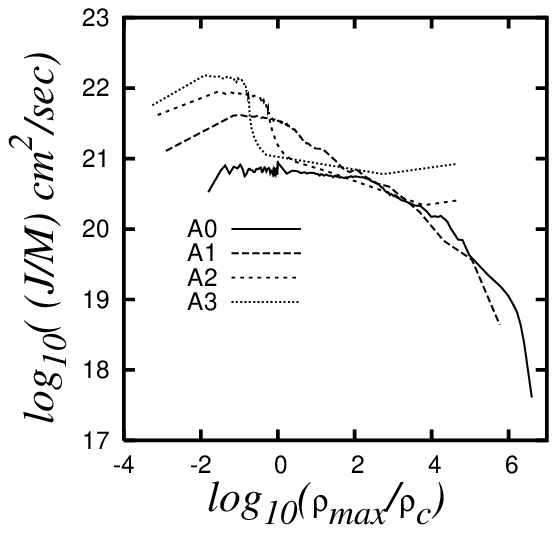}&\includegraphics[width=3.0in]{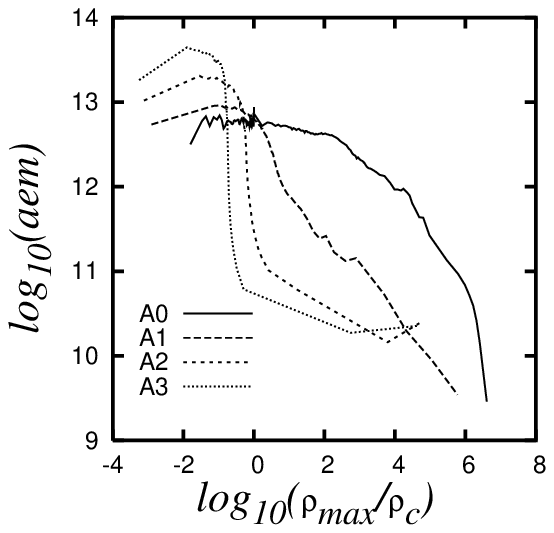}
\end{tabular}
\caption{Distribution of specific angular momentum (left)
and $aem$ ratio (right) against the density
for every particle in the last snapshot available
in each simulation.}
\label{DistLyRho}
\end{center}
\end{figure}

\begin{figure}
\begin{center}
\begin{tabular}{cc}
\includegraphics[width=3.0in]{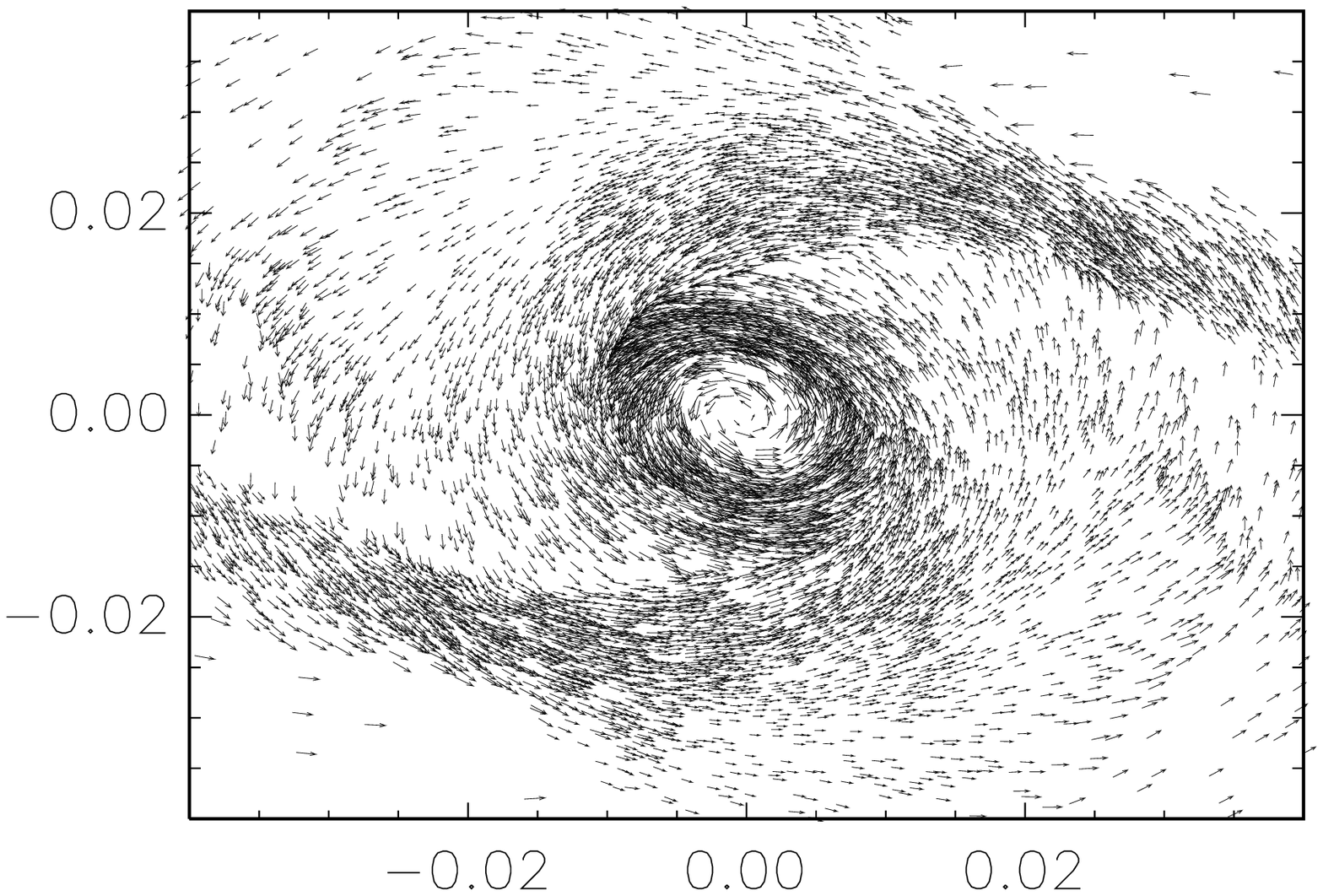}&\includegraphics[width=3.0in]{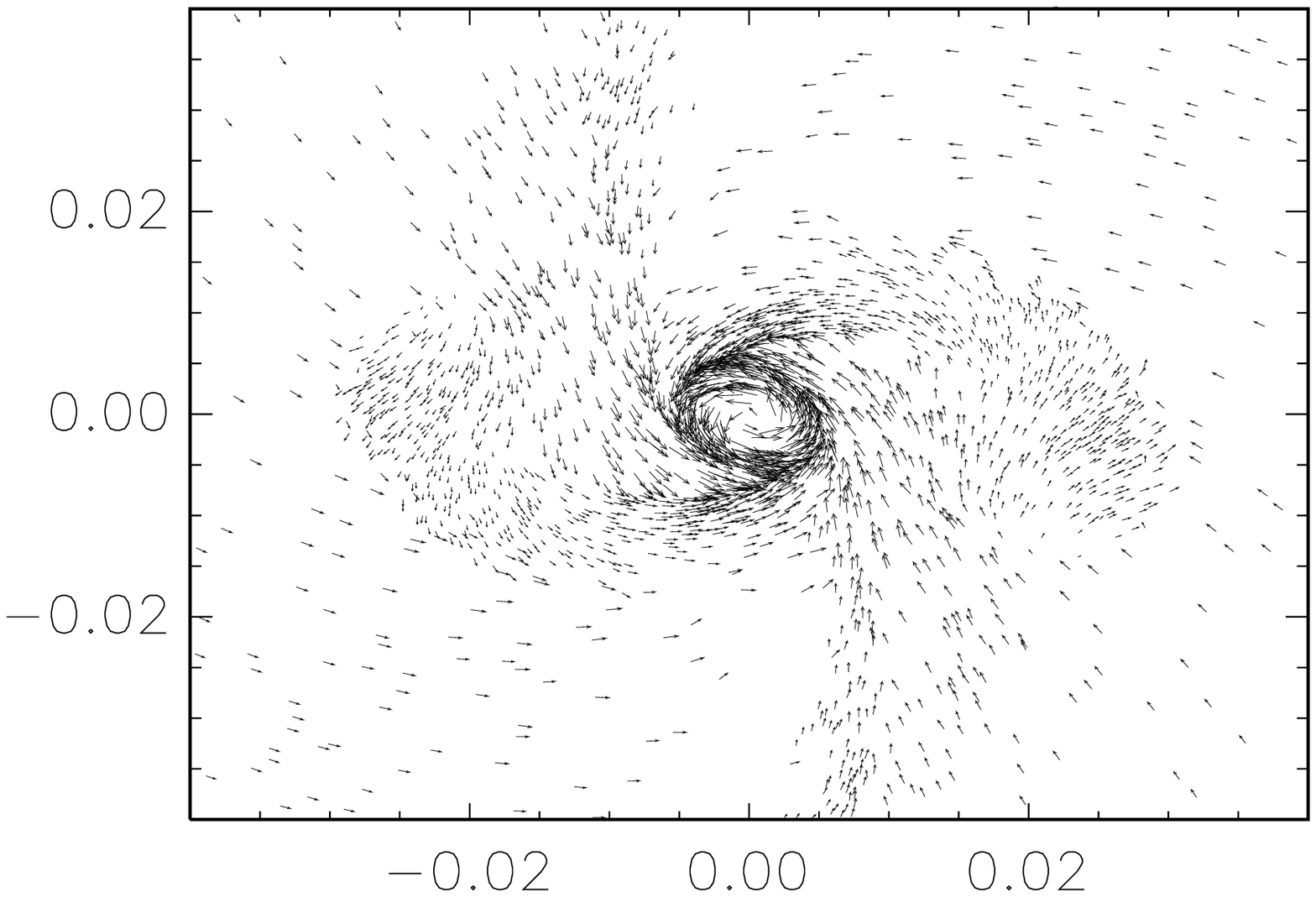} \\
\includegraphics[width=3.0in]{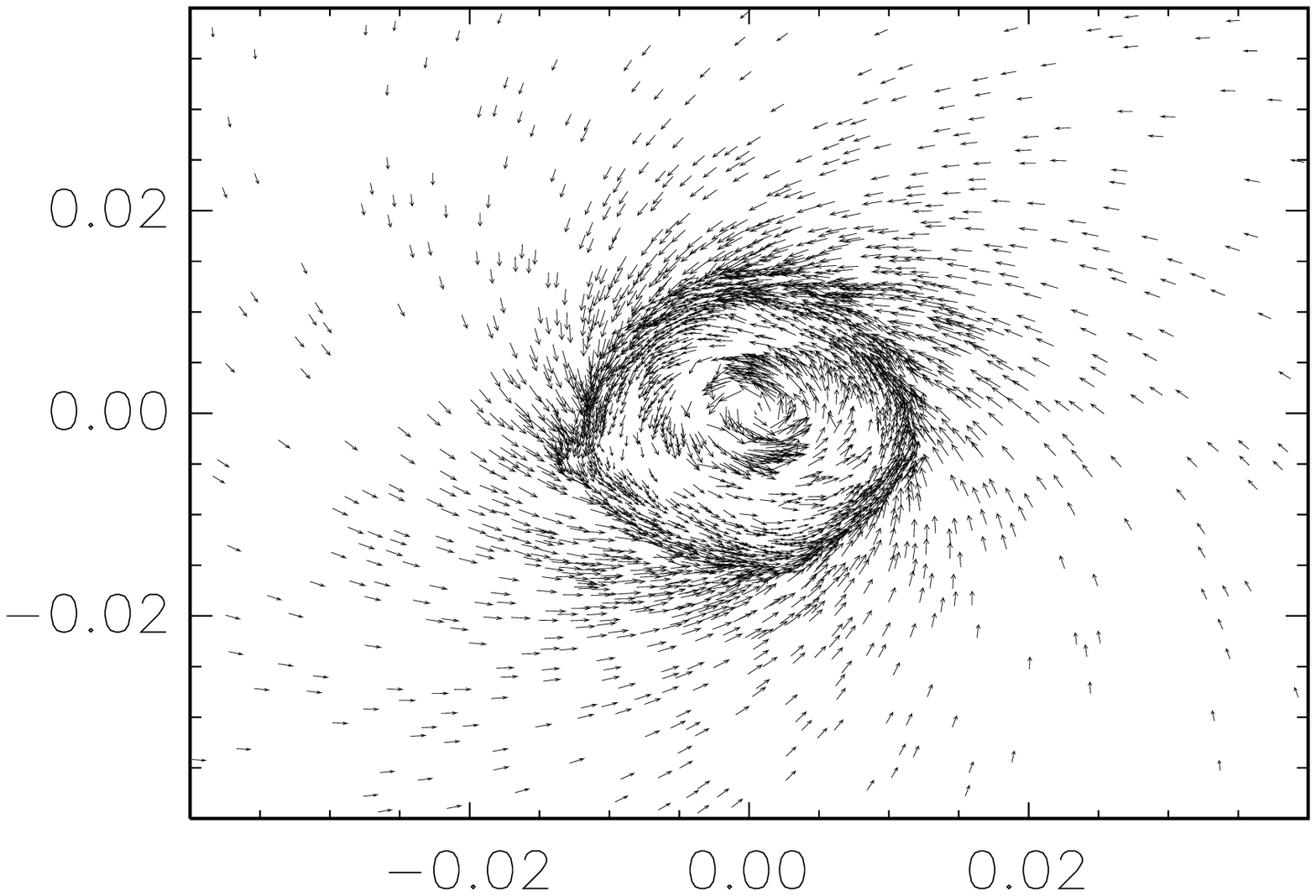}&\includegraphics[width=3.0in]{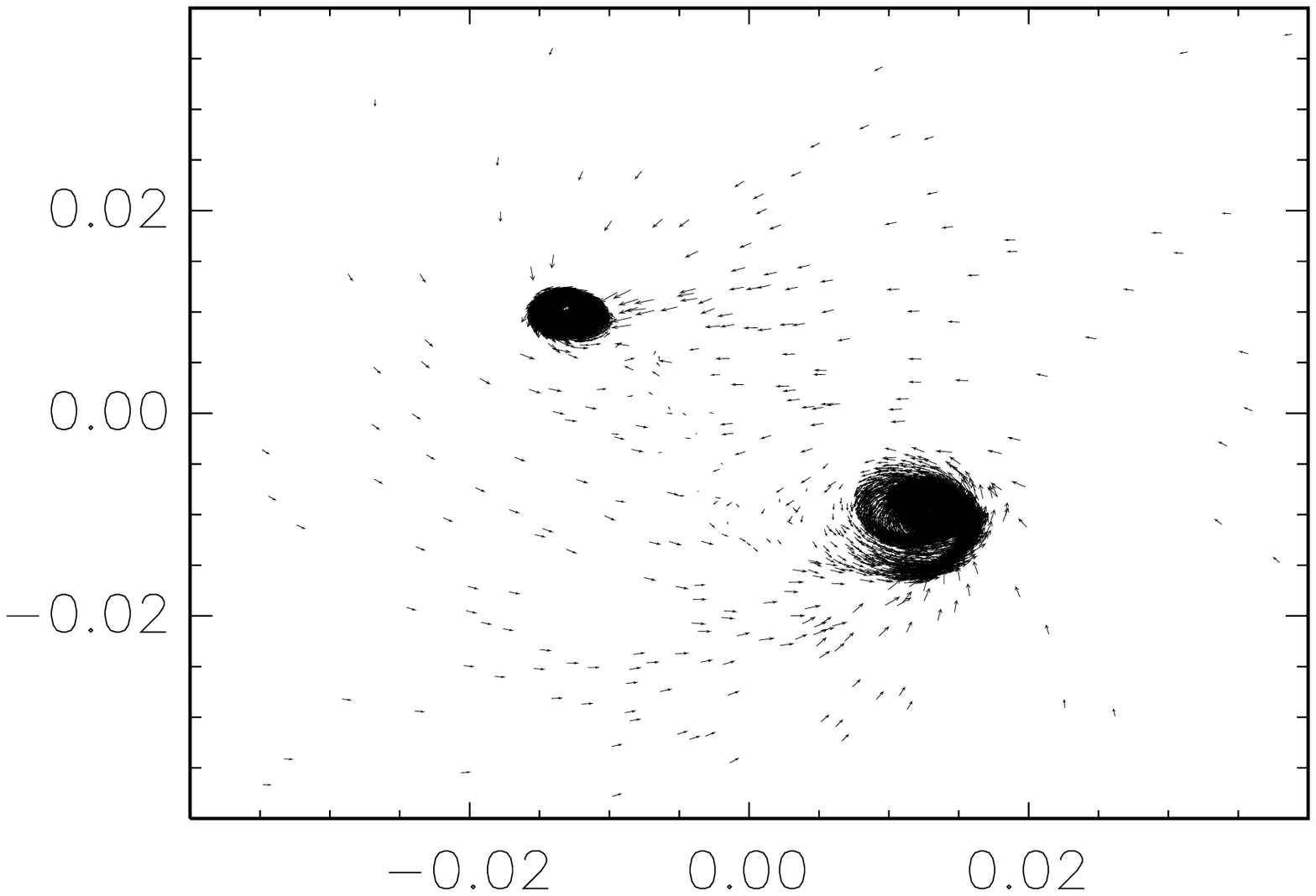}
\end{tabular}
\caption{ A 2D plot with the velocity distribution for the final
snapshot obtained for each simulation,
for model $A0$ (upper left);
for model $A1$ (upper right);
for model $A2$ (lower left) and
for model $A3$ (lower right). The axes in all of these panels
are $x/R_c$ and $y/R_c$, as is the case in Fig.~\ref{SeguiDer}.}
\label{VelDistri}
\end{center}
\end{figure}

\begin{figure}
\begin{center}
\begin{tabular}{cc}
\includegraphics[width=3.0in]{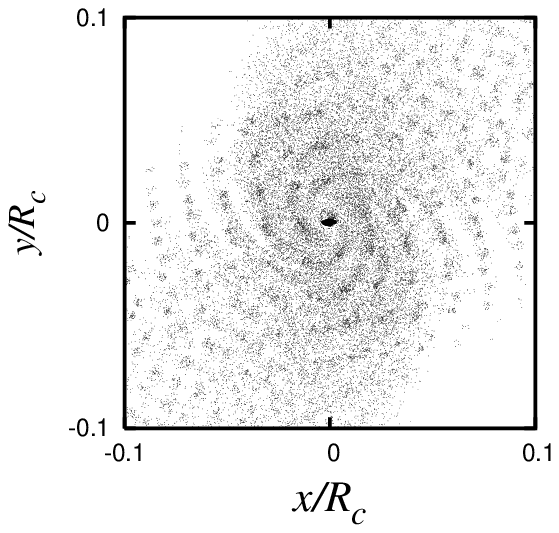}&\includegraphics[width=3.0in]{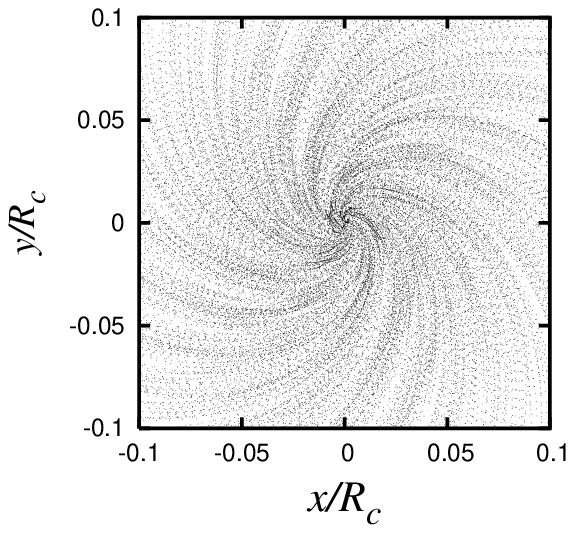} \\
\includegraphics[width=3.0in]{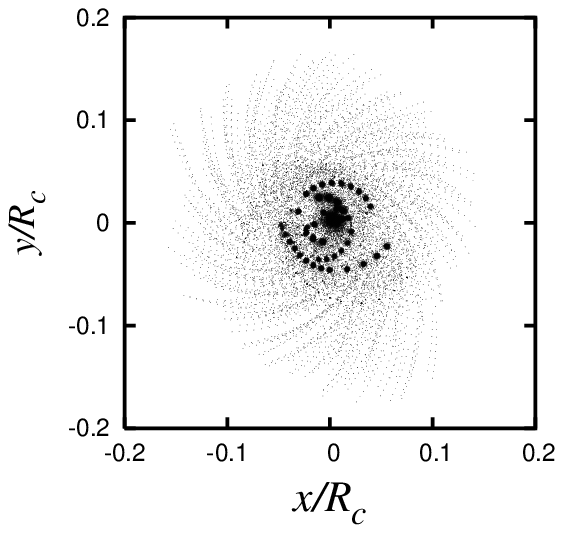}& \includegraphics[width=3.0in]{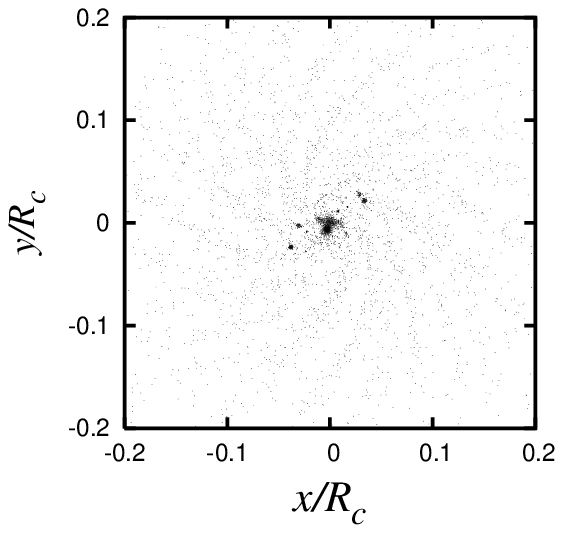}
\end{tabular}
\caption{A 2D view of the path followed by a given set of particles being accreted
by the central cloud region, where the densest clumps are forming. Each dot in these
plots represents a $SPH$ particle of the simulation. The panels here are
displayed in the same order that in Fig.~\ref{VelDistri}.}
\label{SeguiDer}
\end{center}
\end{figure}

\begin{figure}
\begin{center}
\begin{tabular}{cc}
\includegraphics[width=3.0in]{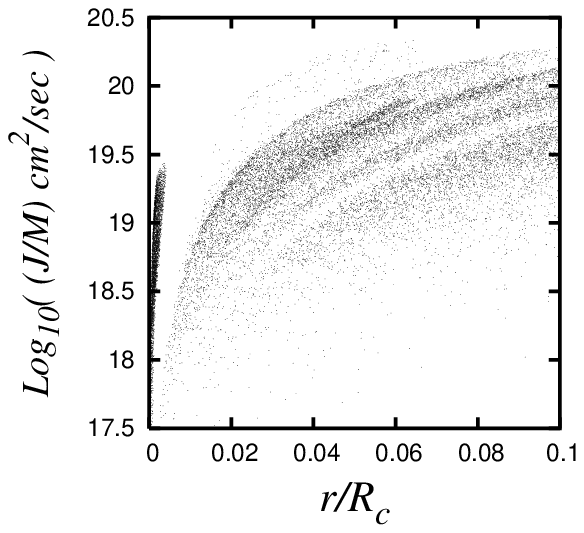}&\includegraphics[width=3.0in]{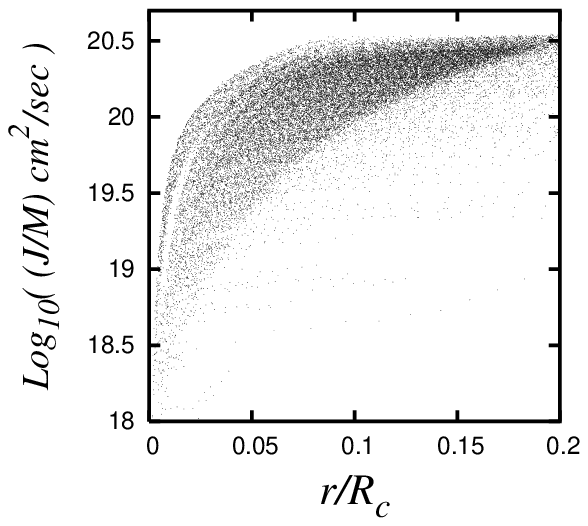} \\
\includegraphics[width=3.0in]{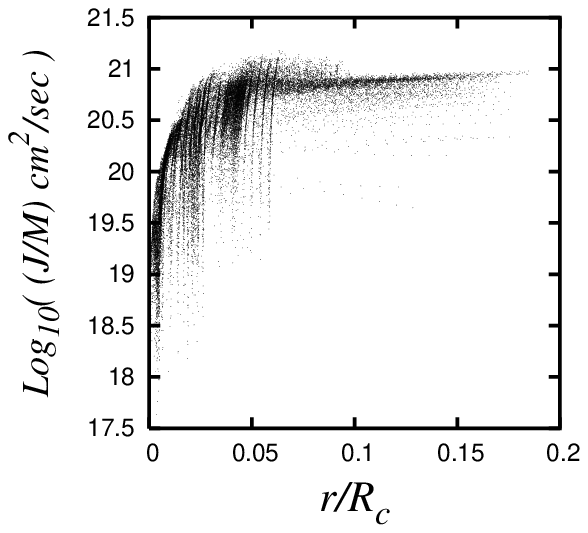}& \includegraphics[width=3.0in]{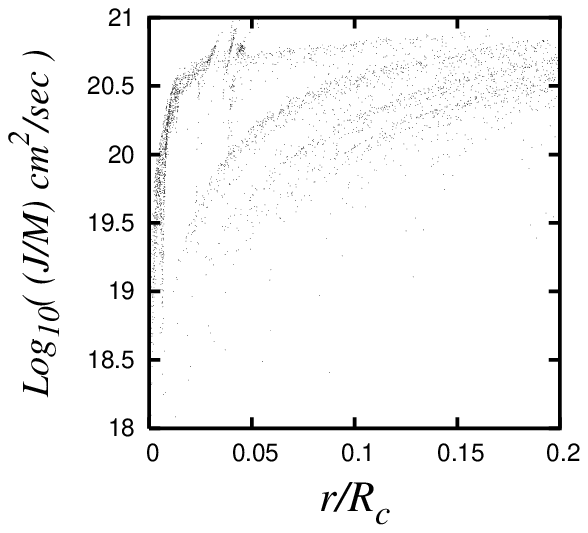}
\end{tabular}
\caption{The specific angular momentum against the radial location for all
the particles of the set already shown in Fig.~\ref{SeguiDer}. The panels here
are displayed in the same order that in Fig.~\ref{VelDistri}. }
\label{SeguiDerLvsR}
\end{center}
\end{figure}

\begin{figure}
\begin{center}
\begin{tabular}{cc}
\includegraphics[width=3.0in]{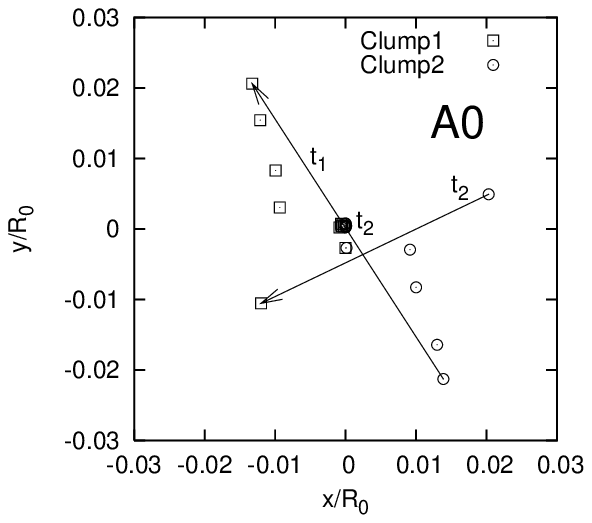}& \includegraphics[width=3.0in]{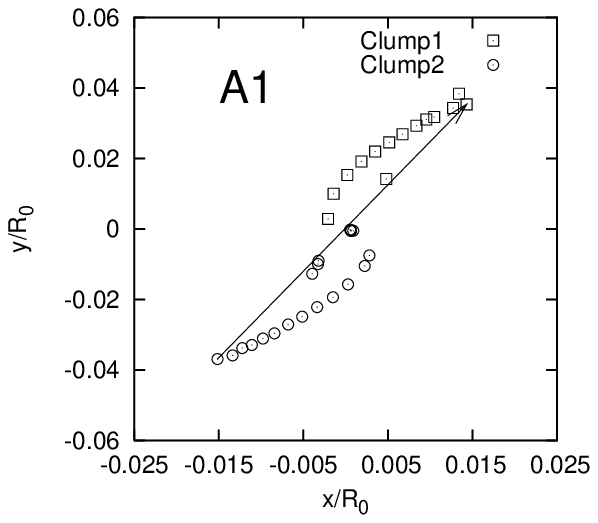}\\
\includegraphics[width=3.0in]{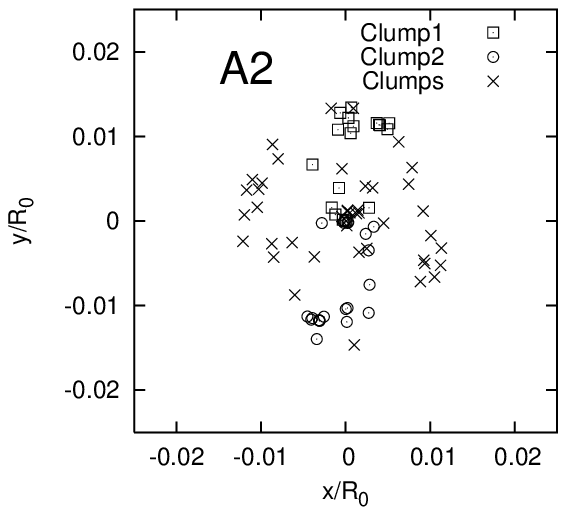}& \includegraphics[width=3.0in]{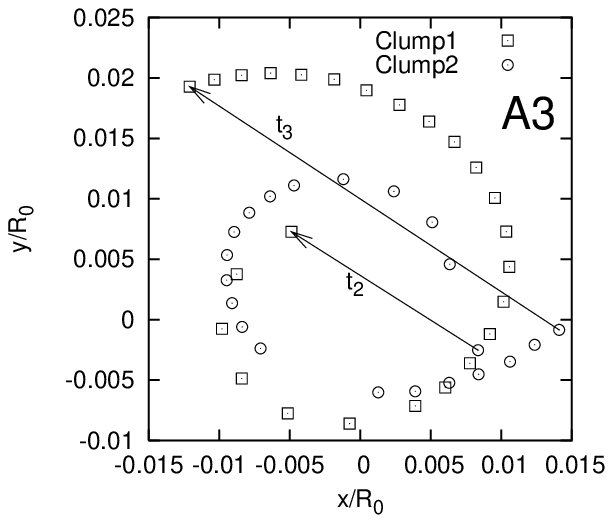}
\end{tabular}
\caption{The path of the centers of clumps already identified
for the cloud models. The lines and the
time labels attached to them, indicate pairs of 
fragments observed at the same time. For a given simulation, we always
have $t_1<t_2<t_3$.}
\label{CentrosForModels}
\end{center}
\end{figure}

\begin{figure}
\begin{center}
\begin{tabular}{cc}
\includegraphics[width=3.0in]{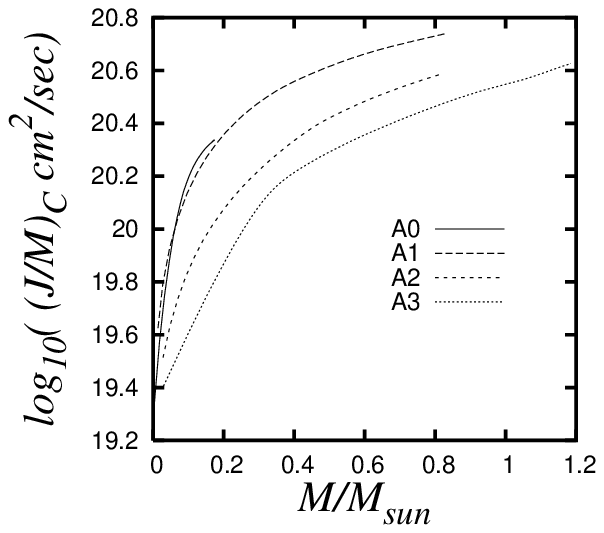}&\includegraphics[width=3.0in]{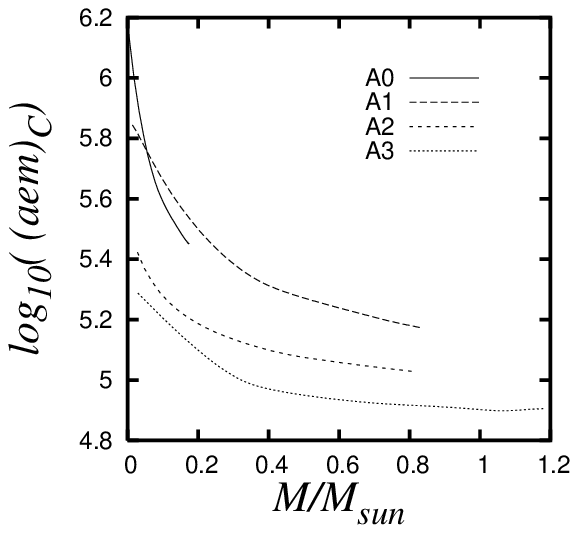}
\end{tabular}
\caption{The angular momentum of the identified clump against the
mass of the same clump, including all those particles with a density higher
than $\rho_{min}=1.4 \times 10^{-17} \; gr/cm^3$ for all models.}
\label{AngMomVSMasalrhomin}
\end{center}
\end{figure}

\begin{figure}
\begin{center}
\begin{tabular}{cc}
\includegraphics[width=3.0in]{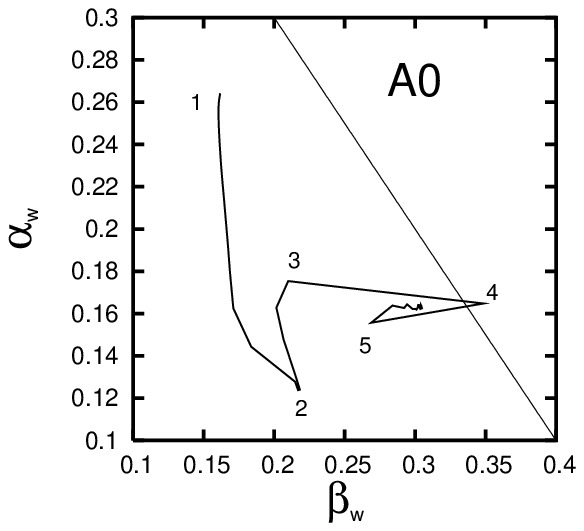}&\includegraphics[width=3.0in]{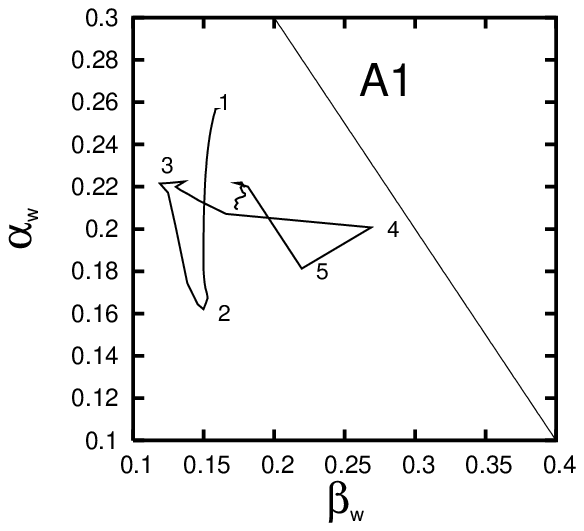} \\
\includegraphics[width=3.0in]{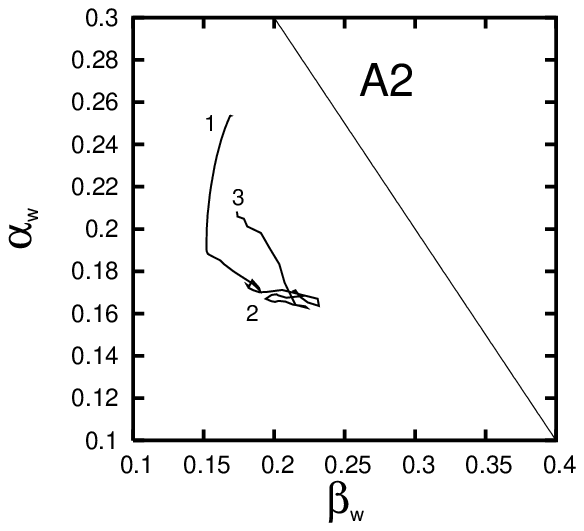}& \includegraphics[width=3.0in]{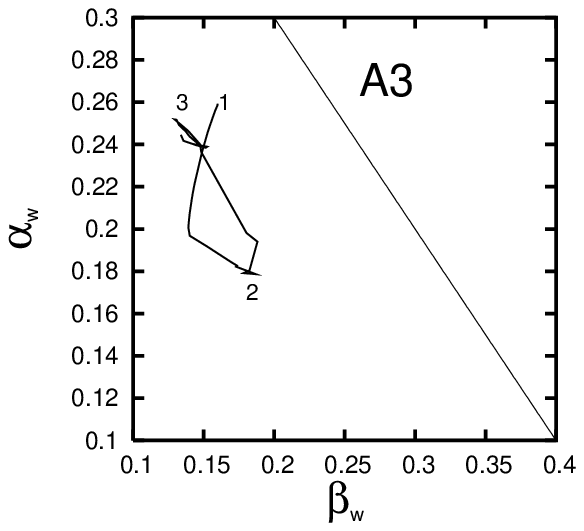}
\end{tabular}
\caption{Energy ratios calculated including all the particles in each simulation.
The virial line is shown as a diagonal and continuous line. See Eq.~\ref{abvirial}.}
\label{IntPropSnap}
\end{center}
\end{figure}

Let us now define with the labels 2-3, a second stage in the
$\alpha$ {\it vs} $\beta$ curves evolution, in which there is a pronounced increase
of the $\alpha$ values. At this stage, which will also occur on all the
models, new clumps are starting to form out. We notice indeed the appearance of
two small over-dense clumps in each extreme of the prolate cloud central region,
which were planted by means of Eq.~\ref{pertmasa}. Shortly after, we noticed that
these small clumps get connected by a very well defined bridge of
particles. As an instance, for the first pairs of models $A0$ and $A1$,
this 2-3 stage has been illustrated with the second and third panels
of Fig.~\ref{Mos4A0} and the second panel of Fig.~\ref{Mos4A1}, respectively.

Later on, we notice that the formation of large spiral arms surrounding the
central clump in models $A0$ and $A1$, has the consequence of an increase in
the $\beta$ value, giving place to a third stage in the $\alpha$ {\it vs} $\beta$
curves evolution, labeled as 3-4 in Fig.~\ref{IntPropSnap}. As the mass of the
central clump increases, the centripetal force acting on the gas should also increase
for those particles with a small radii, pointing out to a strong increase in the
rotational energy.

We observe in Fig.~\ref{IntPropSnap} that the 3-4 stage does not occur neither
in model $A2$ nor in $A3$, despite of the fact that we observe that spiral arms are
also formed in these models, although with a much smaller extension that in
models $A0$ and $A1$.

There is still a 4 stage for the pairs of models $A0$ and $A1$, labeled
by 4-5, in which both the $\alpha$ and the $\beta$ values decrease, indicating that
the cloud central region is losing both thermal and rotational energy. The main
dynamic event occurring at this stage is the merger of the two clumps already
formed, as illustrated in the bottom right panel of
Fig.~\ref{Mos4A0} for model $A0$, and the bottom left panel of
Fig.~\ref{Mos4A1} for model $A1$.

We finish the dynamic description of models $A0$ and $A1$ by noticing that
there is a last stage of the curve after the point 5 label, in which we observed the
formation of {\it exterior} clumps resulting from the spiral arms breakage.

There is also a last stage for models $A2$ and $A3$ manifested in the rattle 
behavior of the $\alpha$ {\it vs} $\beta$ curve, just after label 2, where
we see the cloud fragmentation by means of tiny clumps being formed around the
central original clump. For instance, in model $A2$, a gas ring surrounding
a central clump formed at the end of the first stage (see the fourth panel of
Fig.~\ref{Mos4A2}) now begins to fragment, as it can be better appreciated in
Fig.9 of \citet{NuestroAA}. We also recognize the cloud central region
fragmentation occurrence for model $A3$, as it can be seen in the last
panel of Fig.~\ref{Mos4A3}.

\section{Discussion}
\label{sec:dis}

We have tried in the Paper hereby to make a link between
a simulation and the behavior of its
associated dynamic variables; particularly, with the angular momentum, the
$aem$ ratio and with the $\alpha$ and $\beta$ parameters. Mainly in
Section ~\ref{subs:GlobalProp} we have shown the way in which the most
important dynamic events of each simulation considered here are manifested 
and recorded in the $\alpha$ and $\beta$ curves displayed in Fig.~\ref{IntPropSnap}.

We have established pairs of simulations due to the similarities
recognized in their $\alpha$ vs $\beta$ curves, pointing out
strong dynamical similarities between the models $A0$ with $A1$
and between the models $A2$ with $A3$. However, there are significant dynamical
differences even among a single pair, which are noteworthy.

We consider now in this Section some characteristic events of each simulation, and
show how these events can be a consequence of either the gas envelope
extension or the simulation initial conditions. We would like to emphasize
here how these events are caused by (or manifested on) the dynamical variables
describing the cloud evolution.

\subsection{The merging issue of the early densest clumps}
\label{subsubs:themergingissue}

The merging process described in Section \ref{subs:GlobalProp} is
a very important dynamical characteristic observed for the pair of
models $A0$ and $A1$, where two
clumps, each formed near the filament end, merge into one
single central matter clump.

As we have implemented a symmetric mass perturbation with respect to the
origin of coordinates of the cloud equatorial plane, the seed clumps that
will form will be antipodes of each other, in such a manner
that an imaginary line joining them will pass through the coordinates
origin, too. Thus, every clump exerts a gravitational torque on the other
clump. The particles velocity in either clump begins to align with the
imaginary symmetry axis joining the clumps, that is, $\vec{v} \approx \vec{r}$,
with the net effect that these particles lose angular momentum. Next, the
particles which are accreted into the clumps are the the ones with lower angular
momentum, whereas those accreted into the surrounding spiral arms are those
with higher angular momentum. As the clumps lose their total angular momentum, the
gravity force that every clump exerts on the others brings them closer
until they finally merge.

The benchmark of uniform density cloud collapse is 
the development of a gas filament with a small gas clump located at 
each filament end. Indeed, \citet{NuestroRMAA} have considered  
a model (labeled as UA), which is very similar to the present 
model A0. It is observed in these models that the 
filament becomes shorter in time due to the gravitational attraction 
between the small gas clumps at its ending points. The models outcome 
are different at the final 
evolution stage, for we observed the formation of a binary system 
in model UA and only one central clump in model A0. We may also  
mention another simulation reported 
by \citet{bate95}, in which the closeness among the clumps was observed, 
but not the final merging. In those works, 
the clumps pass by each other without merging, settling
into an orbit around each other. 

The reason behind the different behavior, which may
decide whether merging occurs or not, is perhaps due to the existence of small
variations in the particles positions and velocities, coming 
from the randomness of the initial particle distribution,
whose origin is the density perturbation in
Eq. \ref{pertmasa}, which turns out to play a very important role in this merging
issue. Indeed, the mass perturbation is the cause for the two small
regions development.

Once the cloud has acquired a flattened configuration, the higher density gas
will form an elliptical structure with these small accretion
regions, at the focal points. As demonstrated by \citet{bh04}, the fate of
this elliptical structure is always to collapse into a filament with a
strong mass accumulation at each ending point.
\subsection{The development of spiral arms}
\label{subsubs:thespiralissue}

The differences in the spiral arms development are in the extension and in the
spiral arms breakage level, as it can be appreciated by looking at
Fig.~\ref{VelDistri}. We have observed the formation of  very large and massive
spiral arms for model $A0$. We still see the formation of  well defined spiral
arms for model $A1$, but now the particles circular motion is somewhat distorted
at the spiral end regions, which may be a warning sign of the forthcoming
breakage. The spiral arms for models $A2$ and $A3$ are thin and short, but
still well defined.

Curiously, the spiral arms are able to fulfill a complete turn around the
central clump, in models $A2$ and $A3$, indicating  that there are more
particles being accreted as well onto the gas ring formed from the spiral arms
rather than directly onto the central clump. We then find that the gas ring has
a very short life term in the case of model $A3$, while the lifetime is longer for
model $A2$.

The key for understanding such different behavior in our simulations, comes from
Section~\ref{sec:inicialener}. It was shown there that the total rotational
energy of the assembled cloud must be shared between the core and the
envelope. We reported in Table~\ref{tab:alphasybetascores} that the rotational
energy remaining in the core was decreasing as the gas envelope mass was
growing. We claim that this core rotational energy is responsible for the
cloud spiral arms growth; consequently, that it
was ultimately the reason behind the different outcomes in the simulations,
whereas for the pairs of models $A0$ and $A1$, the rotational energy left
in the core is enough for the spiral arms formation, while for the pair of
models $A2$ and $A3$, it is not.
\subsection{The fragments virialization issue}
\label{subsubs:thevirissue}

We showed in Section~\ref{subs:GlobalProp} that when we include
all the simulation particles in the calculation of the energy ratios
$\alpha$ and $\beta$, the curves do not show any trend to approach
the virial line. We would conclude in this case that neither the resulting
fragments nor the cloud itself virialize, and that the cloud
collapse is still in progress. We emphasize that we have not followed the
subsequent simulation time evolution because the time-step of the run becomes extremely
small, to the point of being almost incapable of advancing the simulation particles
forward in time.

However, when we calculated the $\alpha$ and $\beta$ values taking
into account only those particles satisfying the cutting parameters,
as defined in Section~\ref{subs:propclumpsandfragments},
we observe that the clumps do show a clear tendency to virialize, as it can be
appreciated in Table~\ref{tab:PropIntFrag}. We emphasize that a similar
conclusion can be drawn from the calculation of ~\citet{NuestroRMAA}, where plots
of the $\alpha$ and $\beta$ time evolution were presented.
\section{Conclusions.}
\label{sec:con}

We carried out in this paper a full set of three dimensional numerical hydrodynamical
simulations, in order to theoretically study the sensitivity of the gas core
gravitational collapse on the extension of a gas envelope surrounding the
core at a high spatial resolution, with a barotropic equation of state and within
the framework of the $SPH$ technique. We have also used several dynamical
variables aiming to characterize the simulation results and the collapsing process
itself. What we have observed in this paper can be summarized as follows:

\begin{itemize}

\item A bigger gas envelope delays much longer the collapse; however, the collapse
is more homogeneous as many particles reach higher densities at
the same time, see Fig.~\ref{evolucionmodelos}.

\item The larger the gas envelope extension, the larger the radius of the gas ring
surrounding the central densest clump; a region which shows resistance to the
collapse, due to the combination of both thermal and centrifugal effects;
see Fig.~\ref{aceletotal}.

\item The gas envelope radial extension does not affect the radial profile behavior
of the specific angular momentum and $aem$ ratio as we only observed changes in
the magnitude, but the same trend for all the models; see Fig.~\ref{JenMPerfil}
and \ref{aemPerfil}.

\item For the models with higher gas envelope extension, the particles falling
into the central densest clump are losing angular momentum as well
as $aem$ ratio even though those particles have not reached
densities as high as those in the models with less gas
envelope extension; see ~Fig.\ref{DistLyRho}.

\item The smaller the gas envelope extension, the larger the spiral arms
extension; ~Fig.\ref{VelDistri}.

\item The larger the gas envelope extension, the smaller the
central densest clump spatial extension and, consequently, the longer the
paths followed by the accreting particles; see ~Fig.\ref{SeguiDer}.

\item The larger the gas envelope extension, the smaller the spatial extension
of the central cloud region around the forming densest clump, where a strong
influence on the loss of angular momentum and $aem$ ratio of the accreting
particles, is observed; see ~Fig.\ref{SeguiDerLvsR}.

\item The gas envelope extension length can drastically change the simulation
final outcome; see Fig.~\ref{CentrosForModels}.

\item As the gas envelope extension increases for the forming clump, 
the specific angular momentum also increases, but the $aem$ ratio
decreases; see Fig.~\ref{AngMomVSMasalrhomin}.

\item The $\beta$ ratio maximum values, as well as the $\alpha$ ratio minimum
values, reached during cloud contraction, are both somehow regulated by the gas
envelope extension; see Fig.~\ref{IntPropSnap}.

\end{itemize}

\section*{Acknowledgments}
GA. would like to thank ACARUS-UNISON for the use of their
computing facilities in the making of this paper.

\newpage
\clearpage
\newpage
\clearpage
\end{document}